%% file: IEEE-conference-template-062824.tex
\documentclass[conference]{IEEEtran}
\IEEEoverridecommandlockouts
\usepackage{cite}
\usepackage{amsmath,amssymb,amsfonts}
\usepackage{algorithmic}
\usepackage{graphicx}
\usepackage{hyperref}
\usepackage{subcaption}
\usepackage{textcomp}
\usepackage{xcolor}
\usepackage{pgf,tikz}
\usepackage{multirow}
\usepackage{multicol}
\usepackage{siunitx}
\def\BibTeX{{\rm B\kern-.05em{\sc i\kern-.025em b}\kern-.08em
    T\kern-.1667em\lower.7ex\hbox{E}\kern-.125emX}}
\begin{document}

\title{A Quantum Approach for Optimal Transient Control in Network-Based Epidemic Models
}

\author{\IEEEauthorblockN{Deborah Volpe\IEEEauthorrefmark{1},
Giacomo Orlandi\IEEEauthorrefmark{2}, Mattia Boggio\IEEEauthorrefmark{2},  Carlo Novara\IEEEauthorrefmark{2}, Lorenzo Zino\IEEEauthorrefmark{2}, and Giovanna Turvani\IEEEauthorrefmark{2} }

\IEEEauthorblockA{
\IEEEauthorrefmark{1} Istituto Nazionale di Geofisica e Vulcanologia, Rome, Italy.\\
\IEEEauthorrefmark{2}Department of Electronics and Telecommunications,
Politecnico di Torino
Italy\\
\href{mailto:deborah.volpe@ingv.it}{deborah.volpe@ingv.it}, 
\href{mailto:giacomo.orlandi@polito.it}{giacomo.orlandi@polito.it},
 \href{mailto:mattia.boggio@polito.it}{mattia.boggio@polito.it},\\
\href{mailto:lorenzo.zino@polito.it}{lorenzo.zino@polito.it},
\href{mailto:carlo.novara@polito.it}{carlo.novara@polito.it},
\href{mailto:giovanna.turvani@polito.it}{giovanna.turvani@polito.it}\vspace{-20pt}}} 

\maketitle

\begin{abstract}
Effective epidemic control is crucial for mitigating the spread of infectious diseases, particularly when pharmaceutical interventions such as vaccines or treatments are limited. Non-pharmaceutical strategies, including mobility restrictions, are key in reducing transmission rates but require careful optimization to balance public health benefits and socioeconomic costs. Quantum computing is emerging as a powerful tool for solving complex optimization problems that are intractable for classical methods and can thus be leveraged to handle mobility restrictions.

This article presents  a new approach to optimizing epidemic control strategies using quantum computing techniques. We focus on non-pharmaceutical interventions, particularly mobility restriction, modeled as a discrete-time network epidemic process based on the susceptible–infected–susceptible and susceptible–infected–removed frameworks. The control problem is formulated as a combinatorial optimization task, inherently NP-hard due to the binary nature of intervention decisions. To tackle this computational complexity, we derive a Quadratic Unconstrained Binary Optimization representation of the control problem, enabling its efficient solution via quantum computing resources. Our methodology is validated through numerical simulations on realistic case studies, showcasing the potential of quantum algorithms for enhancing epidemic control strategies. These findings pave the way for leveraging quantum optimization in broader applications of networked dynamical systems, demonstrating its viability for complex decision-making processes in public health management.
\end{abstract}


\section{Introduction}

\begin{figure*}
    \centering
\includegraphics[width=0.8\textwidth]{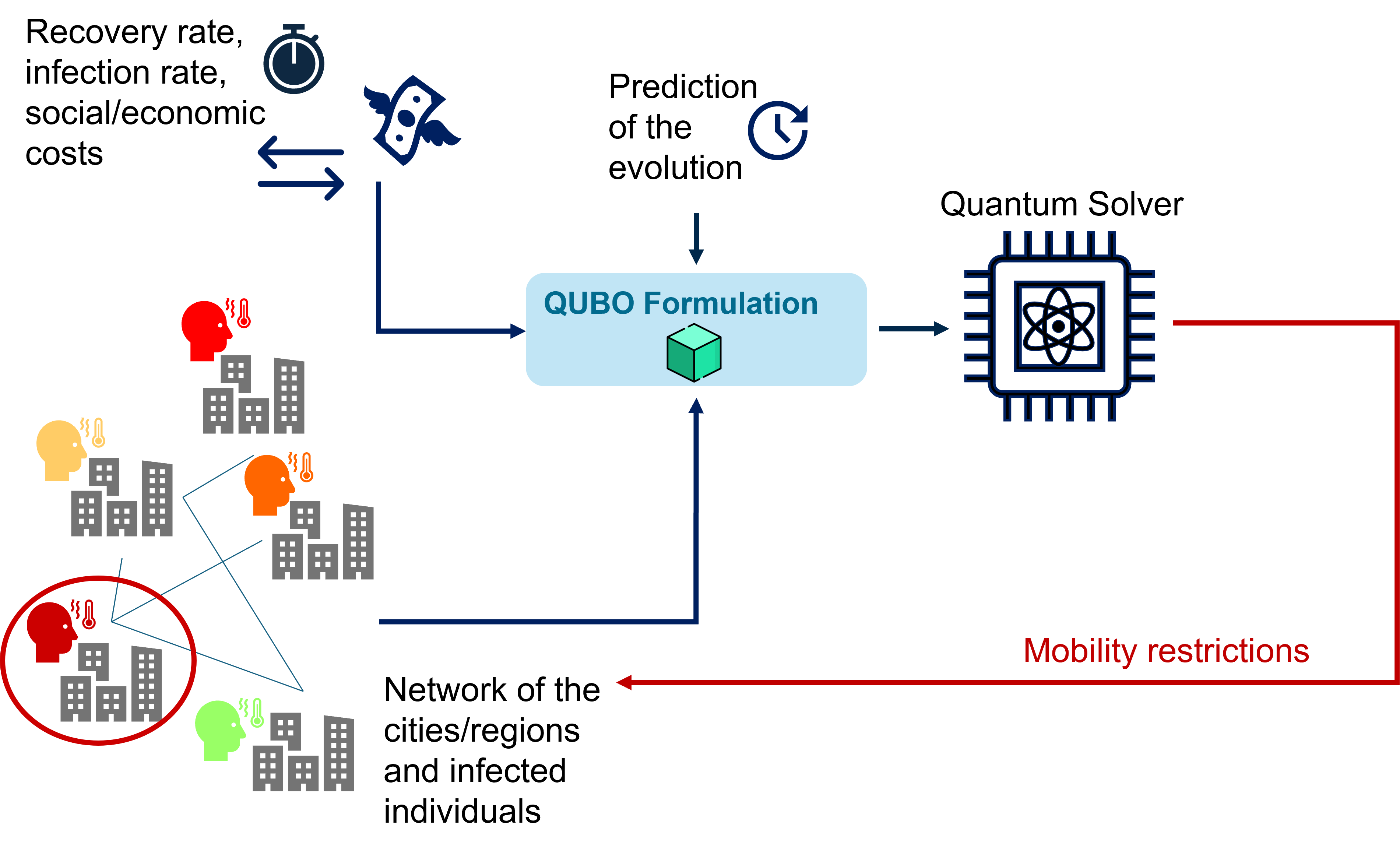}
    \caption{Proposed quantum approach for optimal transient control in network-based epidemic models. }
    \label{fig:GA}
\end{figure*}

The integration of epidemic modeling with network theory plays a crucial role in control strategies aimed at designing and optimizing intervention policies~\cite{Nowzari2016,Holme2016,
Ogura2019,Mei2017,Pare2020review,zino2021survey,Humphries2021,Ye2023competitive}. Among the many challenges posed by an epidemic outbreak, a critical issue is the design of effective intervention strategies when pharmaceutical solutions, such as vaccines or antiviral treatments, are not available. For example, during the COVID-19 pandemic, many countries implemented strict mobility restrictions, including nationwide lockdowns, travel bans, and localized containment zones~\cite{Iezadi2021,Dergiades2022}. The primary objective of these interventions is to flatten the epidemic curve while simultaneously accounting for the socioeconomic consequences of these measures, ensuring that public health decisions are effective and sustainable~\cite{Fenichel2013,Ash2022}. Network-based epidemic models can help policymakers determine the most appropriate interventions, assess their impact, and strike a balance between controlling disease spread and maintaining economic stability~\cite{DellaRossa2020,Carli2020,Giordano2021,Khler2021,parino2021rs,Ye2021,Hota2021,Cenedese2021,Acemoglu2021,Frieswijk2023,Galante2024,She2024,Zino2024cdc}. 

However, controlling network epidemic models through \textbf{non-pharmaceutical interventions} (\textbf{NPIs}) is a computationally expensive problem. Deciding to restrict mobility between two different regions can be represented as a binary variable, and finding the optimal set of mobility restrictions translates into a combinatorial optimization problem. Therefore, the mathematical formulation of epidemic control is network-based and involves binary variables with at most quadratic relationships. Consequently, such optimization problems naturally fit within the \textbf{Quadratic Unconstrained Binary Optimization} (\textbf{QUBO}) framework~\cite{Volpe2025}. This model describes the combinatorial optimization problem using a quadratic pseudo-boolean cost function, representing the problem as follows:
\begin{equation}
    \textrm{minimize} \quad f(\mathbf{x}) = q_0 + \sum_i q_ix_i + \sum_{i<j} q_{ij}x_ix_j \, ,
\end{equation}
where $\mathbf{x}$ is the vector of binary variables, $q_i$ and $q_{ij}$ are respectively linear and coupling coefficients, and $q_0$ is a constant term that can be neglected and does not affect the solution of the problem. The QUBO model is inherently NP-hard~\cite{barahona1982complexity}; however, it allows solving these problems with quantum computing, as it is the best-suited formulation for \textbf{Quantum Annealing} (\textbf{QA}) \cite{kadowaki1998quantum, ManufacturedSpins,Volpe2025} and quantum gate-based computing algorithms, such as \textbf{Grover Adaptive Search} (\textbf{GAS}) \cite{gilliam2021grover,giuffrida2022engineering} and \textbf{Quantum Approximate Optimization Algorithm} (\textbf{QAOA}) \cite{blekos2024review}. Promising results in the use of quantum computing for solving complex and real-world problems formulated in the QUBO framework can be found in \cite{marchesin2023,volpe2024towards,volpe2024predictive, marchioli2024scheduling,gagliardi2025,novara2025}.

Previous research has explored solving epidemic control problems with classical optimization methods. Yet, many of them focus on the eradication of diseases rather than managing the infection curve following an outbreak~\cite{Drakopoulos2014,Preciado2014,Nowzari2015,Mai2018,Zino2020,Zino2020lcss,Somers2021,Walsh2025}. In addition, with increasing mobility both between regions and across nations, the scale of these problems can grow significantly when considering larger geographic areas. Finer-grained control measures further expand the problem size, as mobility restrictions must be carefully handled to balance public health benefits with their economic impact.

In this article, we consider two fundamental network epidemic models: the \textbf{susceptible–infected–susceptible} (\textbf{SIS}) and the \textbf{susceptible–infected–removed} (\textbf{SIR}) models~\cite{zino2021survey}, which differ in whether recovery from the disease provides natural immunity or not. 
We apply a control strategy formulated as an optimization task, balancing outbreak containment with the economic costs of node isolation. We derive a quadratic objective function from the network epidemic models, leveraging cost function aggregation to achieve an optimal trade-off between minimizing infection rates and limiting the number of isolated nodes to sustain economic activity. We analyze various network sizes to assess the scalability of the proposed approach. Additionally, we validate the models using real COVID-19 data and explore different levels of network granularity, accounting for administrative divisions at two hierarchical levels—regions and provinces—in Italy, using real mobility data. We compare the solutions obtained using classical optimization methods with those from D-Wave’s quantum annealers. We also select smaller groups of Italian regions to perform simulations using the QAOA. Our results highlight the benefits of applying quantum computing techniques to real-world optimization problems, establishing a promising synergy with network epidemic control. As quantum technology continues to evolve, it holds the potential to further accelerate computations and efficiently tackle larger problem instances.

The rest of the paper is organized as follows. Section~\ref{sec:motivation} discusses the motivation for using quantum optimization in epidemic control problems. Section~\ref{sec:model} introduces the epidemic model, while Section~\ref{sec:qubo} details the QUBO formulation developed to represent the control problem. In Section~\ref{sec:results}, we present the results obtained, and in Section~\ref{sec:discussion} we discuss the findings and implications. Finally, Section~\ref{sec:conclusions} concludes the paper and outlines potential directions for future research.

\section{Motivation and General Idea}\label{sec:motivation}

The recent COVID-19 pandemic underscored the critical role of \textbf{mathematical modeling} and \textbf{control-theoretic approaches} in planning \textbf{effective intervention policies} to mitigate the spread of infectious diseases. Among these, \textbf{network epidemic models} have proven valuable in catching the complex and heterogeneous transmission dynamics occurring across different geographic areas. These models have been extensively employed to optimize resource allocation, design vaccination strategies, and guide behavioral responses to epidemic episodes.

When pharmaceutical interventions such as vaccines or antiviral treatments are unavailable, NPIs become the primary means of preventing disease spread. One of the most effective NPIs is \textbf{mobility restrictions}, which limit movement between locations to reduce transmission rates. However, implementing these restrictions involves balancing healthcare system capability against social and economic costs, making defining optimal intervention policies a \textbf{challenging optimization problem}.

Mathematically, the problem of optimally deploying mobility restrictions leads to a \textbf{discrete, combinatorial optimization problem} that is \textbf{NP-hard}, making it computationally infeasible to solve with standard classical methods, especially for large-scale networks. Quantum computing has recently emerged as a promising paradigm for handling such problems, as it offers significant speedups for certain combinatorial optimizations by exploiting quantum parallelism and quantum annealing techniques.

In this work, we propose a quantum computing-based approach for optimizing mobility restriction policies in network epidemic models (Figure~\ref{fig:GA}). Specifically, we focus on the SIS and SIR models, which capture the spread of non-immunizing and immunizing diseases, respectively. We describe the problem of determining optimal mobility restrictions as a \textbf{QUBO} problem, a mathematical formulation well-suited for quantum algorithms.

Our main contributions are as follows:
\begin{itemize}
    \item We derive a \textbf{QUBO formulation} for the optimal mobility restriction problem, making it compliant to quantum solvers.
    \item We demonstrate the feasibility of using quantum computing techniques to efficiently solve this problem, exploiting both \textbf{quantum annealing and quantum circuit model paradigms}.
    \item We validate our approach through numerical simulations on \textbf{COVID-19 data}, proving its potential to provide practical decision-support tools for policymakers.
\end{itemize}

This study represents an important step toward exploiting quantum computing for the control of networked epidemic processes, providing a scalable and efficient framework that can be extended to broader applications in public health, urban mobility planning, and complex network dynamics.

\begin{table}
\centering
\caption{Model and optimization parameters.}\label{tab:parameters}
\begin{tabular}{r| l}
$n_i$& population in location $i\in\mathcal L$\\
$\bar x_i$& infected individuals in location $i$ at time $t=0$\\
$\bar y_i$& removed (immune) individuals in location $i$ at $t=0$\\
$A_{ij}$&interactions between population in location $i$ and $j$\\
$\mu$& recovery rate\\
$\lambda$& infection rate\\
$\gamma$&weight of social/economic costs in decision-making
\end{tabular}
\end{table} 

\section{Epidemic Model}\label{sec:model}

We model a population distributed across $M$ locations, denoted as $\mathcal{L} = \{1, \dots, M\}$, where each location $i$ hosts $n_i$ individuals. For large populations, we approximate $n_i$ as a continuous variable ($n_i \in \mathbb{R}_{>0}$). Individuals interact both within and across locations, represented by a weighted graph $\mathcal{G} = (\mathcal{L}, A)$, where the adjacency matrix $A \in \mathbb{R}_{\geq 0}^{M\times M}$ quantifies the intensity of interactions between locations. Typically, $A_{ij} \in [0,1]$, and self-loops are excluded ($A_{ii} = 0$). A schematic representation of this network is shown in Fig.~\ref{fig:network}, and a list of variables and parameters included in the models is reported in Table \ref{tab:parameters}.

\begin{figure}
    \centering
    \input{img/network}
    \caption{Network population model with the entries of weight matrix $A$. }
    \label{fig:network}
\end{figure}
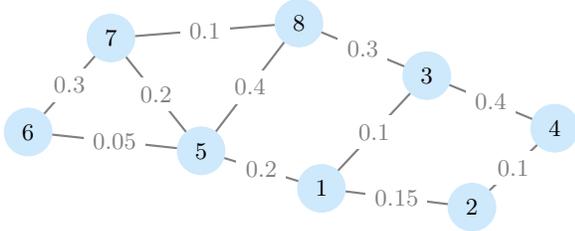

\begin{figure}
    \centering
  \subfloat[SIS model \vspace{2pt}]{\input{img/sis}  \label{fig:sis}}\quad\subfloat[SIR model]{\input{img/sir}  \vspace{-15pt} \label{fig:sir}}\caption{Schematic of the SIS and SIR epidemic models. }\label{fig:schematic}
\end{figure}
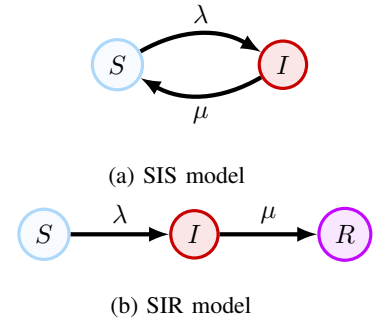

We consider two fundamental epidemic models: the \textbf{SIS} and the \textbf{SIR}, explicitly discussed in~\cite{Mei2017}. Both models categorize individuals as susceptible (S) or infected (I), but differ in post-recovery behavior: 
\begin{itemize}
    \item In the \textbf{SIS} model, recovered individuals immediately re-enter the susceptible state, reproducing diseases like many STIs \cite{Yorke1978}.
    \item In the \textbf{SIR} model, recovered individuals acquire permanent immunity (or have long-lasting immunity relative to the epidemic timescale), as seen in diseases like COVID-19.
\end{itemize}

Each location $i$ has an infected population $x_i(t)$, and in the SIS, the susceptible population is $ n_i - x_i(t)$. The system state is represented by the vector $\mathbf{x}(t) = [x_1(t), \dots, x_M(t)]^\top$, evolving as:
\begin{equation}\label{eq:sis}
    x_i(t+1)=(1-\mu)x_i(t)+\frac{\lambda}{n_i}\left(n_i-x_i(t)\right)\alpha_i(t)\, ,
\end{equation}
where $\lambda$ is the infection rate, i.e. the susceptibility of an individual to become infected,  $\mu$ is the fraction of the infected individual recover, and the infection force  $\alpha_i(t)$ is:
\begin{equation}\label{eq:alpha}
    \alpha_i(t)=x_i(t)+\sum_{j\in\mathcal L\setminus\{i\}} A_{ij}x_j(t)\, .
\end{equation}
The first term accounts for local infections, while the second captures mobility-driven spread. Unlike classical SIS model on networks \cite{Lajmanovich1976}, we explicitly separate these contributions since control interventions act only on the second term.

For the SIR, an additional state variable $y_i(t)$ handles the removed individuals. The system evolves as:
\begin{equation}\label{eq:sir}
\begin{array}{l}
   x_i(t+1)=(1-\mu)x_i(t)+\frac{\lambda}{n_i}\left(n_i-x_i(t)-y_i(t)\right)\alpha_{i}(t),\\
    y_i(t+1)=y_i(t)+\mu x_i(t) \, .
\end{array}
\end{equation}
where the susceptible population is $n_i - x_i(t) - y_i(t)$, reflecting permanent immunity after recovery.

A key property of the model is that, with an appropriate time step selection, the system dynamics remain well-defined. Additionally, the infection rate $\lambda$ relates to the basic reproduction number via $\lambda \propto R_0 \mu$, where the proportionality constant is the largest eigenvalue of $A + I$ (see \cite{zino2021survey} for more details).

We focus on NPIs, particularly \textbf{mobility restrictions}, which have been widely implemented during COVID-19. Policymakers can isolate locations, limiting mobility to control infection spread.
Control is introduced via a binary vector $\mathbf{u} = [u_1, \dots, u_M]^\top$, where:
\begin{equation}\label{eq:control}
    u_i=\left\{\begin{array}{ll}1&\text{if location $i$ is isolated,}\\
    0&\text{otherwise.}
    \end{array}\right.
\end{equation}
This modifies the infection force in Equation~\eqref{eq:alpha} to:
\begin{equation}\label{eq:alpha_control}
    \alpha_i(\mathbf{u},t) = x_i(\mathbf{u},t) + (1 - u_i) \sum_{j\in\mathcal L\setminus\{i\}} A_{ij}x_j(\mathbf{u},t)\, .
\end{equation}
Thus, if $u_i = 1$, the $i^{\textrm{th}}$ location is isolated, and infections occur only within that location.

Although control actions could be time-dependent, we consider optimization over a fixed time window, allowing implementation within a rolling-horizon control framework.

\section{QUBO formulation}\label{sec:qubo}

In epidemic networks, NPIs involve isolating nodes to prevent the spread of infection between areas with high levels of contagion. At the same time, node isolation carries non-negligible economic impact that must be taken into account. Therefore, we built the following aggregate cost function that aims to balance the two objectives:
\begin{equation}\label{eq:qubo}
    f(\mathbf{x}, \mathbf{u}) = \sum_{i \in \mathcal{L}} \sum_{t=1}^T x_i(t) + \gamma \sum_{i \in \mathcal{L}} n_iu_i\, ,
\end{equation}
where $T \in \mathbb Z_{>0}$ is the time-horizon and $\gamma$ is the aggregation coefficient that weights the importance of the economic contribution with respect to the healthcare aspect. This QUBO formulation holds for both the SIR and SIS models, according to the expression of $x_i(t)$ used in Equation \eqref{eq:qubo}. The first term accounts for the minimization of the number of infected individuals, while the second accounts for reducing the costs associated with NPIs, being proportional to the population of the areas where isolation is enforced. Intuitively, a node is isolated only if its impact on the healthcare system is larger than the economic consequences of isolation. %Here, we consider all nodes to have economic relevance only depending on the population of the area.

The number of infected individuals at time step $t+1$ is affine in the decision variables $\mathbf{u}$, as per Equation~\eqref{eq:alpha_control}. Thus, the time-horizon $T$ determines the degree of the polynomial. In fact, the time-horizon represents the number of recursive steps to calculate the expression of $x(t)$. As a result, in Equation \eqref{eq:alpha_control}, the variable $u_i$ is multiplied by $x_j(\mathbf{u}, t)$, which in turn recursively depends on the same product. Therefore, the polynomial includes at most $T$ multiplications involving $u$ variables.

Equation \eqref{eq:qubo} represents the general expression which can be provided to \textbf{PUBO} (\textbf{Polynomial Unconstrained Binary Optimization}) solvers, such as GAS and QAOA. However, from hereinafter we consider $T=2$ to have a QUBO-compliant formulation that is also solvable by quantum annealers. Epidemic network control problems with longer time horizons can still be formulated as QUBO problems by using exact methods to convert polynomial terms into quadratic ones through the introduction of auxiliary variables \cite{anthony2017quadratic}. Alternatively, approximated methods, such as recursively applying Taylor expansion, can be used to reduce the polynomial to second-order terms.

\section{Results}\label{sec:results}

In this section, we present the epidemic trends observed optimizing the network control through node isolation. We compare the simulated epidemic progression, based on both real and synthetic data, across networks controlled by several classical and quantum optimization methods. 

\begin{figure}[h]
    \centering
\includegraphics[width=0.7\columnwidth]{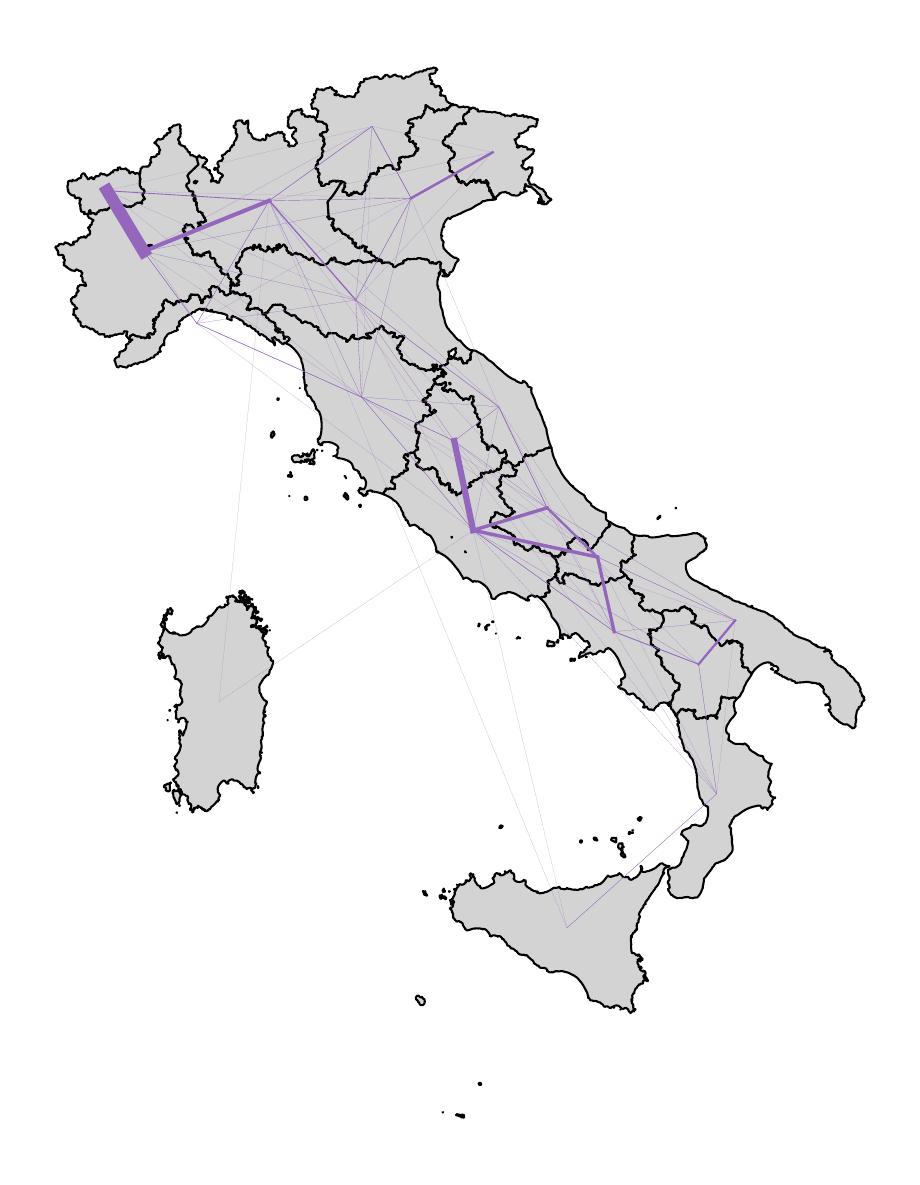}
    \caption{Map of Italy showing the first-level administrative divisions (regions) and the connections between them. The thickness of each link is proportional to the corresponding weight in the adjacency matrix $\mathbf{A}$.}
    \label{fig:Italy}
\end{figure}

\subsection{Considered Scenario}

We evaluated our proposed quantum optimization approach on both a \textbf{synthetic} and a \textbf{real-world epidemic} scenario, the latter calibrated using data from the COVID-19 outbreak in Italy. The analysis spans across different levels of granularity and employs both the SIS and the SIR epidemic models. 

Specifically, for the real-world epidemic data, we consider a network in which each node represents an Italian administrative division, and links (with associated weights) represent individual mobility and travel between these divisions. Using publicly available data from the Italian National Institute of Statistics~\cite{Istat}, we construct networks at different levels of granularity: either with $n=21$ regions (first-level administrative divisions, shown in Figure~\ref{fig:Italy}) or with $n=107$ provinces (second-level administrative divisions). The weight of each link is set proportional to the fraction of the population commuting between two administrative divisions, as reported in~\cite{Istat}. 

For tests involving quantum circuit models, which require smaller networks due to the limitations of current quantum circuit classical simulators, we also consider reduced instances: one involving the provinces within a single region (specifically, Piedmont), and another where the country is coarsely partitioned into three macro-regions --- North, Center, and South --- according to conventional Italian geographic division.

We consider both the SIS and SIR epidemic models, calibrated on the recent COVID-19 pandemic. Accordingly, we set the model parameters following~\cite{parino2021rs}. Simulations are initialized using the officially reported number of cases in each administrative division on the selected date, which varies across simulations to evaluate the impact of the proposed control techniques at different key stages of the epidemic outbreak~\cite{protezioneCivile}.

For each level of granularity, we select four different initial dates corresponding to distinct phases of the epidemic: (i) March 8, 2020; (ii) March 10, 2020; (iii) October 29, 2020; and (iv) December 20, 2021. The first two dates correspond to early phases of the outbreak, with a moderately low number of reported cases. The third date reflects a scenario of rising cases due to the emergence of a novel variant. The fourth date represents a later stage, in which COVID-19 had become endemic, yet a new wave was approaching due to increased mobility associated with the winter holiday period.

In our simulations, we implement the proposed control technique in a rolling horizon fashion, over a total time span of 50 weeks for the SIS model and 30 weeks for the SIR model. Specifically, starting from the initial simulation date (denoted as $t=0$), we solve the control problem at each time step $t$ using the QUBO formulation presented in Section \ref{sec:qubo}, over a time window of $T=2$. The optimal control policy $\bf{u^*}$ is then applied for a single time step. At time $t+1$, the same procedure is repeated, and this process continues until the end of the simulation horizon is reached.

To evaluate how optimization time scales with problem size, we also generate synthetic data consistent with the characteristics of real-world scenarios.

\subsection{Solvers}

In this work, we aim to assess solution quality and execution times of quantum optimization, formulated as a QUBO problem, as opposed to classical methods. For this purpose, we use and evaluate the following solvers:
\begin{itemize}
    \item \textbf{Simulated annealing} (\textbf{SA}) \cite{henderson2003simulated};
    \item \textbf{Tabu search} (\textbf{TS}) \cite{glover1998tabu};
    \item \textbf{Genetic algorithm} (\textbf{GA}) \cite{frenzel1993genetic};
    \item \textbf{Quantum annealing} (\textbf{QA}) \cite{kadowaki1998quantum, ManufacturedSpins,Volpe2025};
    \item \textbf{Quantum Approximate Optimization Algorithm} (\textbf{QAOA}) \cite{blekos2024review}.
\end{itemize}

SA, TS, and QA are all executed using the \texttt{D-Wave Solver} library (considering default options), while GA is implemented using the \texttt{pymoo} optimization library (with a population size of 20). QAOA is executed using the \texttt{qiskit-optimization} library (considering standard Mixed State Hamiltonian, Hadamard layer for initial state creation and COBYLA optimizer), exploiting a classical simulator of quantum circuits, as current quantum devices do not allow obtaining reliable results. 

These solvers were selected to provide a representative comparison between classical metaheuristics and emerging quantum technologies~\cite{Hussain2018}.

\subsection{Figures of Merit}

To evaluate the effectiveness of the proposed approach, we consider the following figures of merit:
\begin{itemize}
    \item \textbf{Peak reduction} ($p$): the percentage reduction in the maximum number of infected individuals with respect to the uncontrolled dynamics;
    \item \textbf{Average infection reduction} ($a$): the  percentage reduction in the mean number of infected individuals per day  with respect to the uncontrolled dynamics:
    \item \textbf{Time} ($t$): the average time for solving a QUBO instance;
    \item \textbf{Final Cost} ($C$): the average value of the objective function with the solver solution.
\end{itemize}

The QA solving time is obtained by extracting the effective Quantum Processing Unit (QPU) time with the method available in the library. The QAOA solving time is not considered in the evaluation, as the classical simulation of quantum circuit model solvers is computationally expensive and does not reflect the execution time on a real quantum computer. 

\subsection{Performed Tests}
\begin{figure*}
    \centering
\subfloat[Regions, 2020/03/08]{\includegraphics[width=.5\columnwidth]{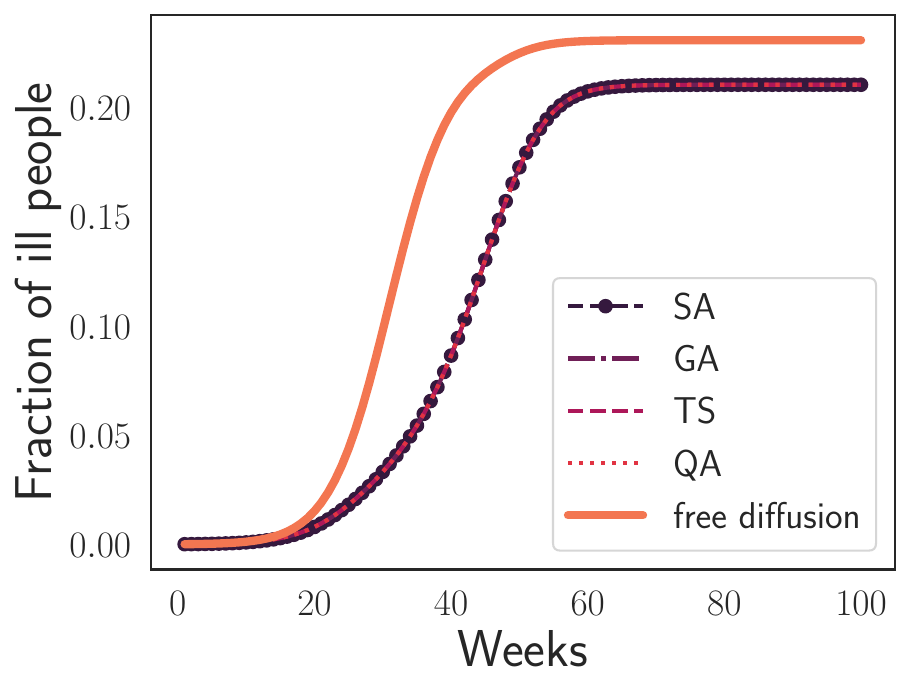}}
\subfloat[Regions, 2020/03/10]{\includegraphics[width=.5\columnwidth]{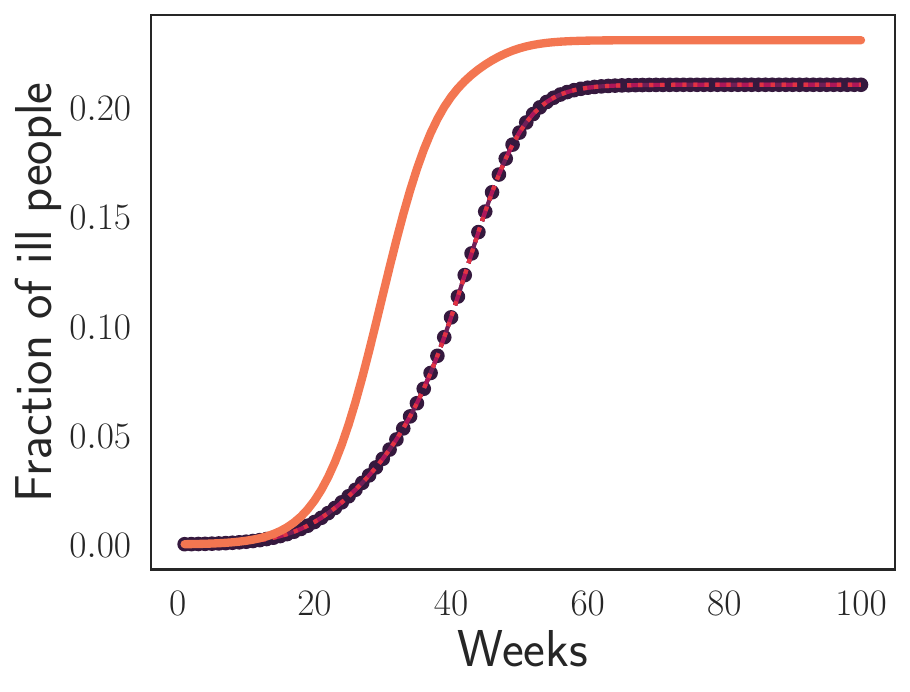}}\subfloat[Regions, 2020/10/29]{\includegraphics[width=.5\columnwidth]{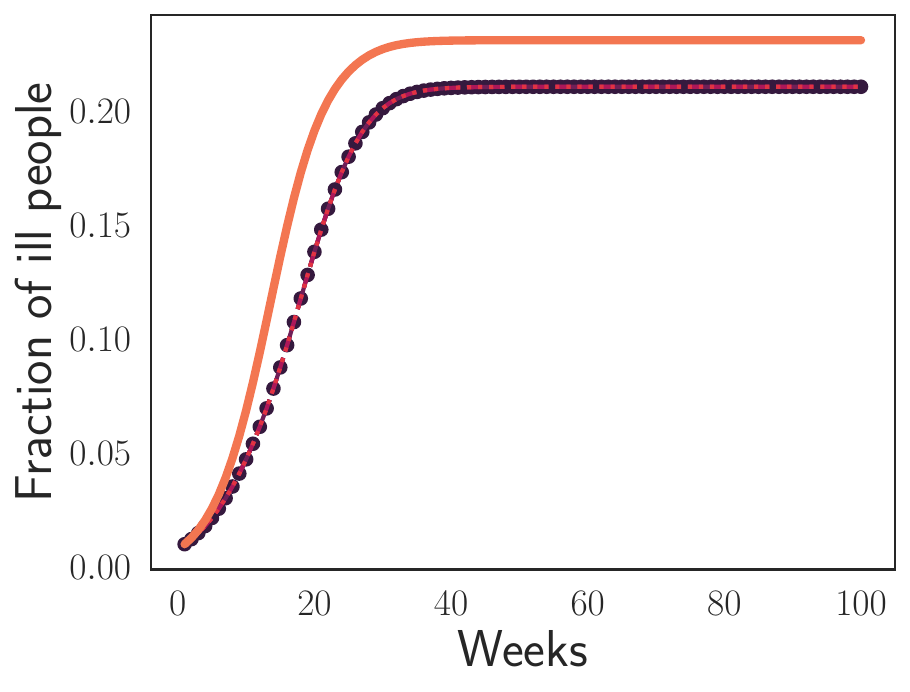}}\subfloat[Regions, 2021/12/20]{\includegraphics[width=.5\columnwidth]{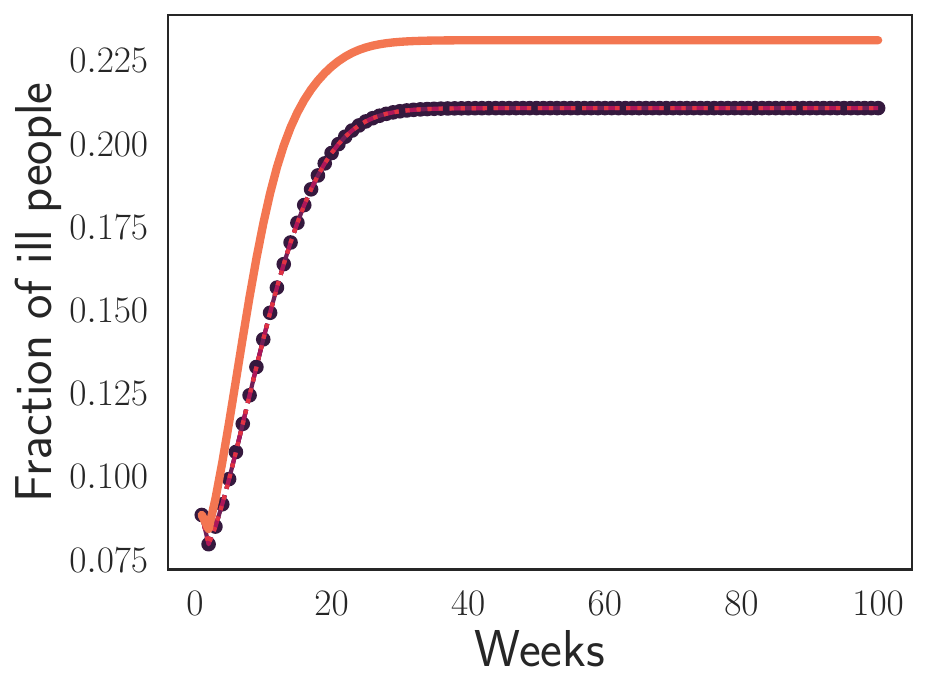}} \\[1ex]
\subfloat[Province, 2020/03/08]{\includegraphics[width=.5\columnwidth]{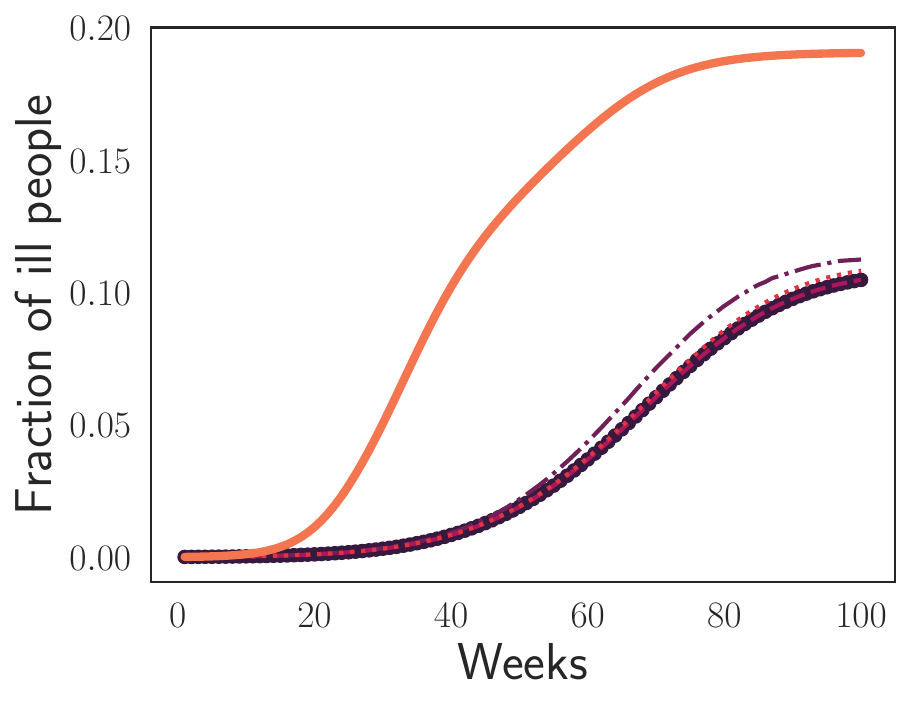}}
\subfloat[Province, 2020/03/10]{\includegraphics[width=.5\columnwidth]{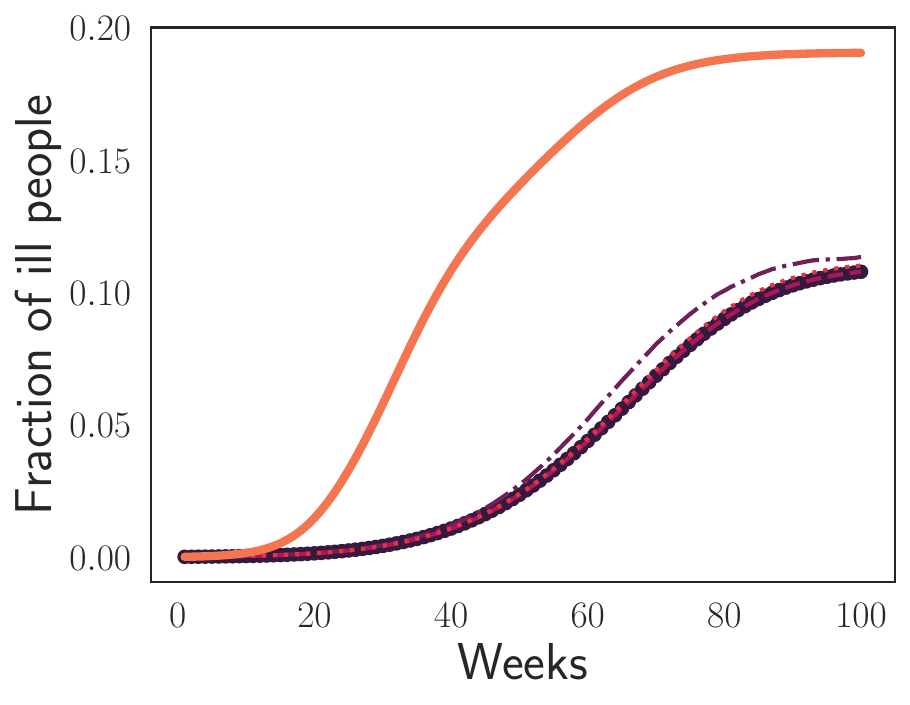}}\subfloat[Province, 2020/10/29]{\includegraphics[width=.5\columnwidth]{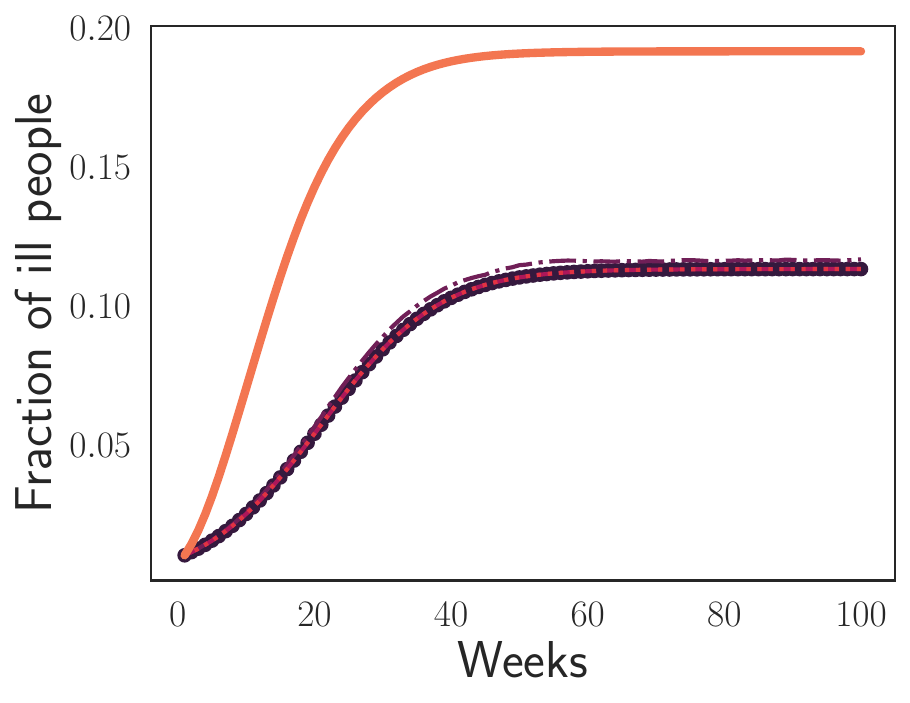}}\subfloat[Province, 2021/12/20]{\includegraphics[width=.5\columnwidth]{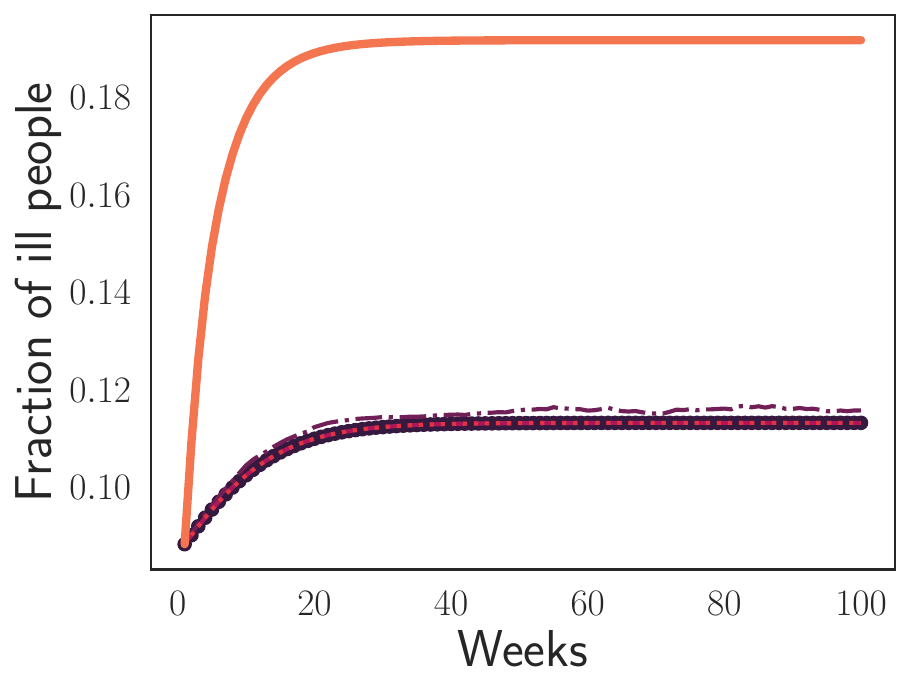}}
    \caption{Results of our numerical simulations (SIS). In (a--d) and (e--h), we report the plots obtained at the granularity of regions and provinces, respectively, with different starting dates. }
    \label{fig:sim}
\end{figure*}
\begin{table*}
    \centering
    \resizebox{0.7\textwidth}{!}{
        \begin{tabular}{|c|c|cc|cc|cc|cc|}
           \hline
            \multirow{2}{*}{\textbf{Data}} & \multirow{2}{*}{\textbf{n}} & \multicolumn{2}{c|}{\textbf{SA}} & \multicolumn{2}{c|}{\textbf{TS}} & \multicolumn{2}{c|}{\textbf{GA}} &  \multicolumn{2}{c|}{\textbf{QA}} \\%& \multicolumn{2}{c|}{\textbf{Free}} \\
            & &  \textbf{p [\%]} & \textbf{a [\%]} &\textbf{p [\%]} & \textbf{a [\%]} & \textbf{p [\%]} & \textbf{a [\%]} & \textbf{p [\%]} & \textbf{a [\%]}%& \textbf{rate} & \textbf{area} 
            \\
            \hline
            2020/03/08 & 21 &  8.83 &  20.95 &  8.83 &  20.95  &  8.83 & 20.95  &  8.83 &  20.95  %& 0.13 & 1.20 
            \\
            2020/03/10 & 21 &  8.83 & 19.75 & 8.83 &19.75 &8.83 & 19.75 & 8.83 & 19.75 %&  0.13 & 1.20
            \\
            2020/10/29 & 21 &  8.83 & 11.73 &  8.83 & 11.73 &   8.83 & 11.73 &   8.83 & 11.73 %& 0.14 & 1.20 
            \\
            2021/12/20 & 21 &  8.83  & 9.93 &    8.83  & 9.93 &    8.83  & 9.93 &   8.83  & 9.93 %& 0.16 & 1.20 
            \\
\hline
            2020/03/08 & 107 & 45.00 & 67.51 &    45.00 &67.51  &  40.95 & 63.19 & 43.17 & 66.51 %& 0.08 & 1.10 
            \\
            2020/03/10 & 107 & 43.40 &  65.04 &  43.40 &  65.04   & 40.46 &  60.94  & 42.15 & 64.14  %& %0.08 & 1.12
            \\
            2020/10/29 & 107 &  40.95 & 45.68 &   40.95 &  45.68 &  39.12 & 43.78 &  40.95 &  45.68 %& 0.13 & 1.16 
            \\
            2021/12/20 & 107 &  40.95 &  40.68 &  40.95 & 40.68 & 39.14 & 39.45 &  40.95 &  40.68 %& 0.22 & 1.18 
            \\
            \hline
        \end{tabular}
    }
    \caption{Performance obtained with different methods in terms of reduction of the peak (p) and average infected/day (a) with respect to the uncontrolled dynamics considering the SIS model. }
    \label{tab:SISPicco}
\end{table*}

\begin{table*}[h]
    \centering
    \resizebox{0.8\textwidth}{!}{ 
         \begin{tabular}{|c|c|cc|cc|cc|cc|}
             \hline
             \multirow{2}{*}{\textbf{Data}} & \multirow{2}{*}{\textbf{N}} & \multicolumn{2}{c|}{\textbf{SA}} & \multicolumn{2}{c|}{\textbf{TS}} & \multicolumn{2}{c|}{\textbf{GA}} &  \multicolumn{2}{c|}{\textbf{QA}} \\
            & & \textbf{t [\si{\milli\second}]} & \textbf{C} & \textbf{t [\si{\milli\second}]} & \textbf{C} & \textbf{t [\si{\milli\second}]} & \textbf{C} & \textbf{t [\si{\milli \second}]} & \textbf{C}\\
             \hline
             2020/03/08 & 21 & 3.40 & 104.31e6 & 2102.92 & 104.31e6 & 87.23 & 104.31e6 & 0.04 & 104.32e6 \\
             2020/03/10 & 21 & 3.39 & 104.94e6 & 2102.83 &  104.94e6 & 88.12 &   104.94e6 & 0.04 &  105.00e6 \\
             2020/10/29 & 21 & 3.12 & 111.30e6 & 2102.57 & 111.30e6 & 79.63 &  111.30e6 & 0.04 & 111.30e6 \\
             2021/12/20 & 21 & 2.15 & 113.63e6 & 2101.42 & 113.63e6  & 48.49 & 113.63e6 & 0.04 & 113.63e6 \\   
\hline
             2020/03/08 & 107 & 1081.021 & 95.35e6 & 2127.60 & 95.35e6 & 533.89 & 81.34e6 & 0.04 & 91.38e6  \\
             2020/03/10 & 107 & 885.14 & 96.63e6 & 2127.97 & 96.63e6 & 512.51 &  82.15e6 & 0.04 & 93.08e6 \\
             2020/10/29 & 107 & 783.09 & 105.00e6 & 2134.16  & 105.00e6 & 598.69 &  91.88e6 & 0.04 & 104.29e6 \\
             2021/12/20 & 107 & 1215.83 & 107.58e6 & 2132.848531 & 107.58e6 & 371.79 & 36.23e6 & 0.04 & 43.37e6 \\     
             \hline
         \end{tabular}}
    \caption{Average solving time and final cost for the different methods for the SIS model.}
    \label{tab:SIS}
\end{table*}
The results obtained at both the regional (21 nodes) and provincial (107 nodes) levels using the SIS epidemic diffusion model are reported in Figure~\ref{fig:sim}, which shows the evolution of the epidemic curve over time under each control policy, compared to the uncontrolled baseline. The plots clearly indicate that all methods significantly reduce the number of infected individuals over time. It is possible to notice that increasing the spatial granularity from regions to provinces results in a further reduction of the total infection rate.

The corresponding quantitative improvements are summarized in Table~\ref{tab:SISPicco}, where it can be observed that QA achieves performance comparable to that of state-of-the-art classical methods in most scenarios. It is worth noting that these results are obtained with a substantially lower computational cost. Table~\ref{tab:SIS} reports the average computation time for each method across all QUBO instances solved throughout the simulations. From this, it is evident that QA offers a remarkable advantage in terms of computational efficiency, operating orders of magnitude faster than classical methods --- particularly in the high-dimensional setting with 107 nodes --- while maintaining comparable solution quality, as confirmed in Table~\ref{tab:SISPicco}. This substantial speed-up makes QA especially attractive for real-time decision-making scenarios, where rapid re-optimization is critical in response to evolving epidemic dynamics.

\begin{figure*}
    \centering
\subfloat[Regions, 2020/03/08]{\includegraphics[width=.5\columnwidth]{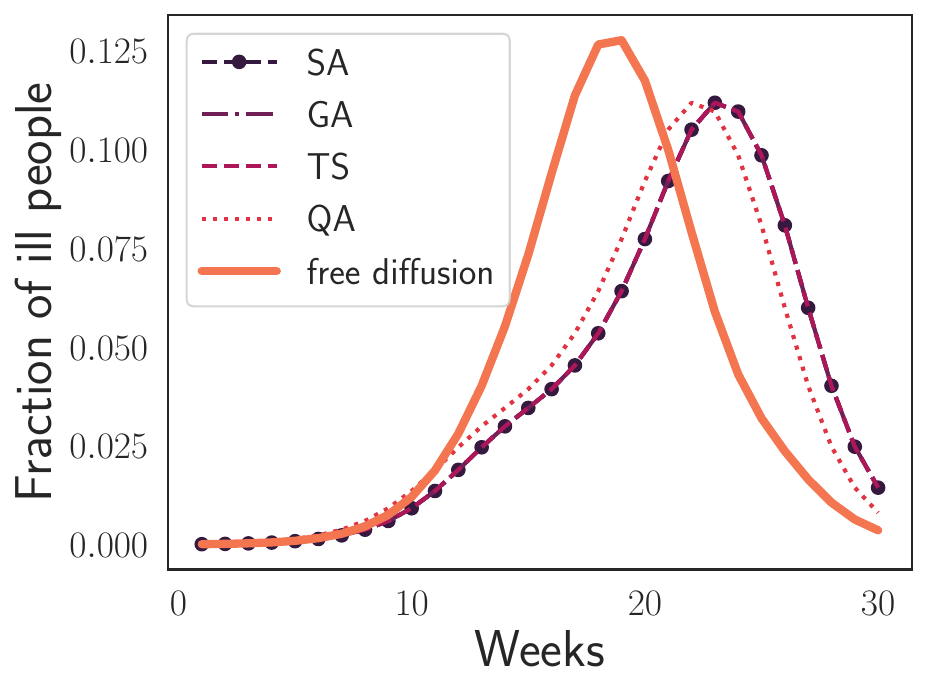}}
\subfloat[Regions, 2020/03/10]{\includegraphics[width=.5\columnwidth]{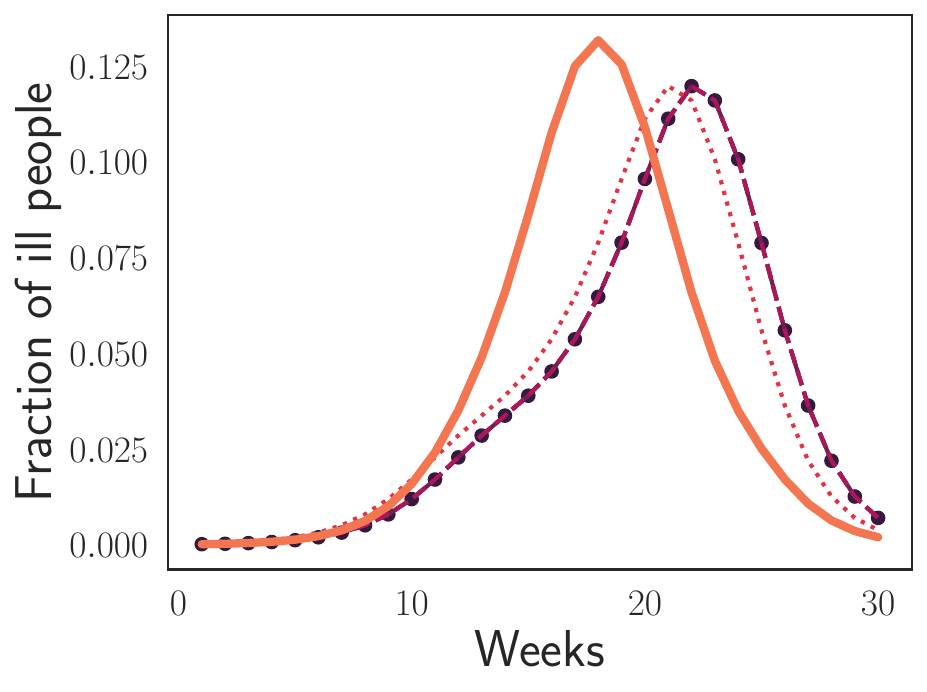}}\subfloat[Regions, 2020/10/29]{\includegraphics[width=.5\columnwidth]{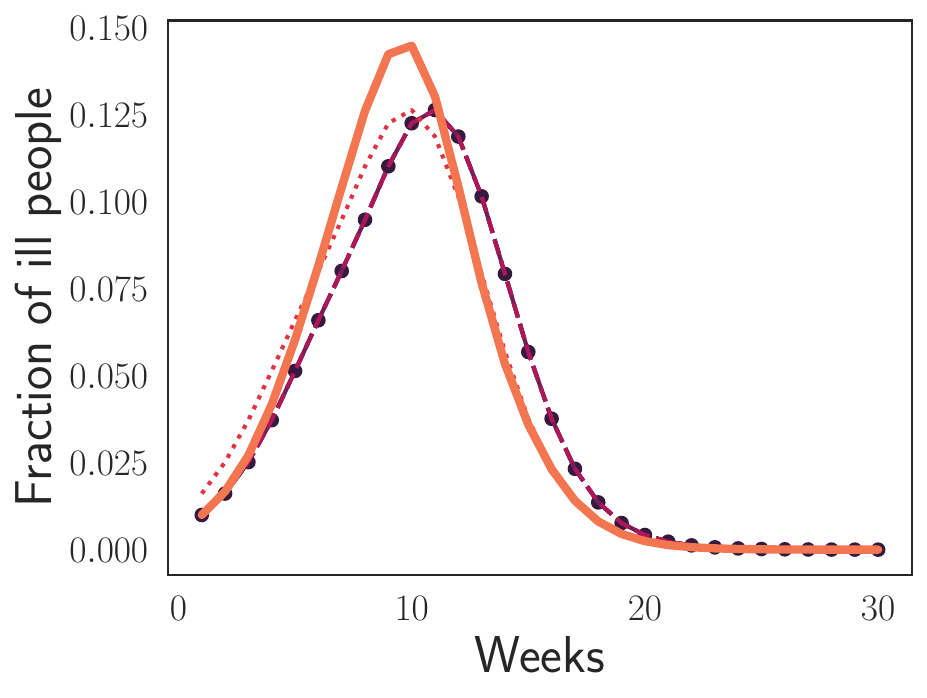}}\subfloat[Regions, 2021/12/20]{\includegraphics[width=.5\columnwidth]{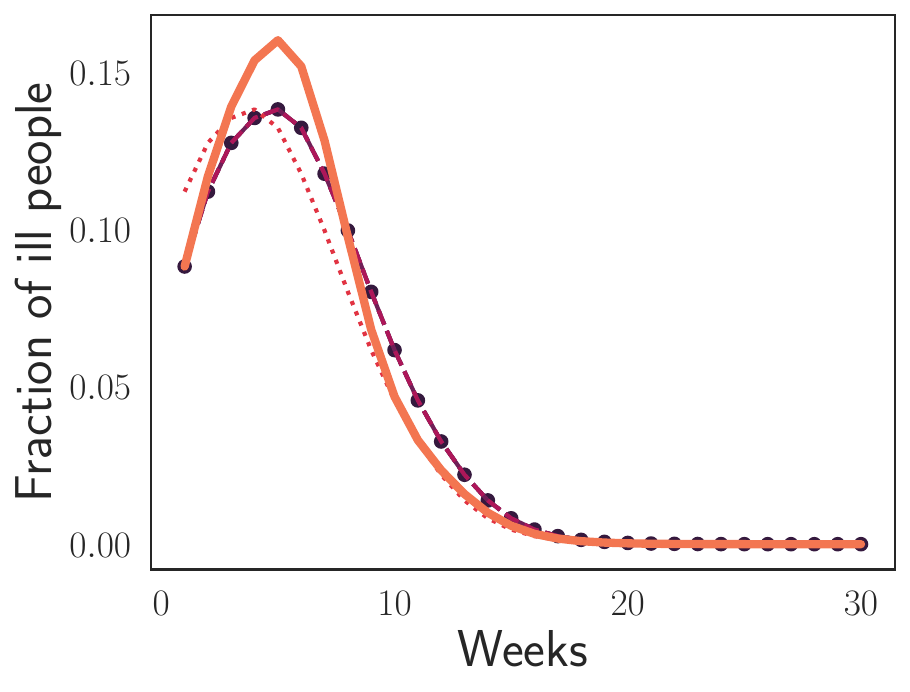}}\\[1ex]\subfloat[Provinces, 2020/03/08]{\includegraphics[width=.5\columnwidth]{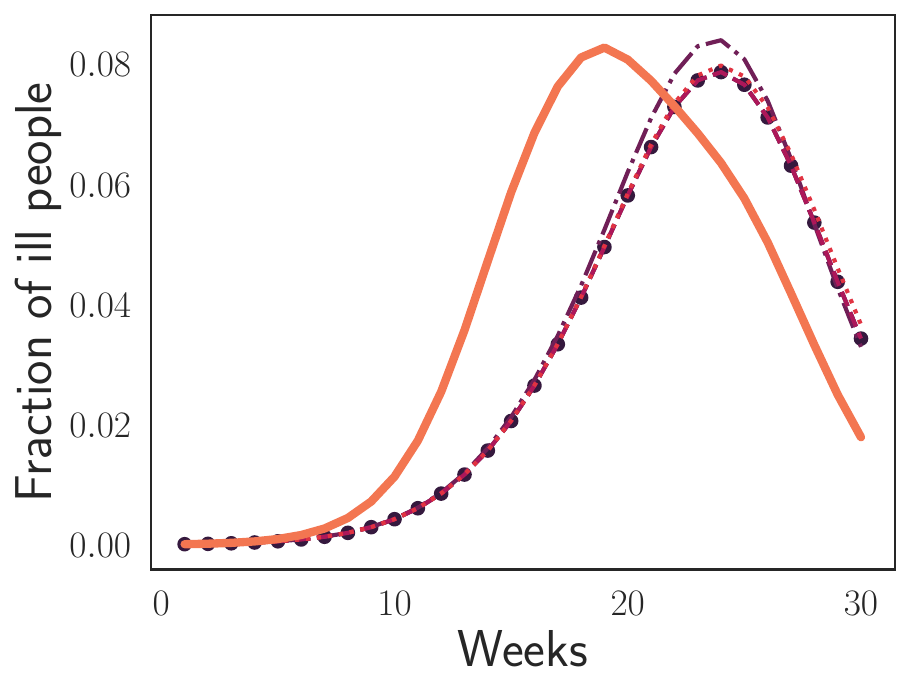}}
\subfloat[Provinces, 2020/03/10]{\includegraphics[width=.5\columnwidth]{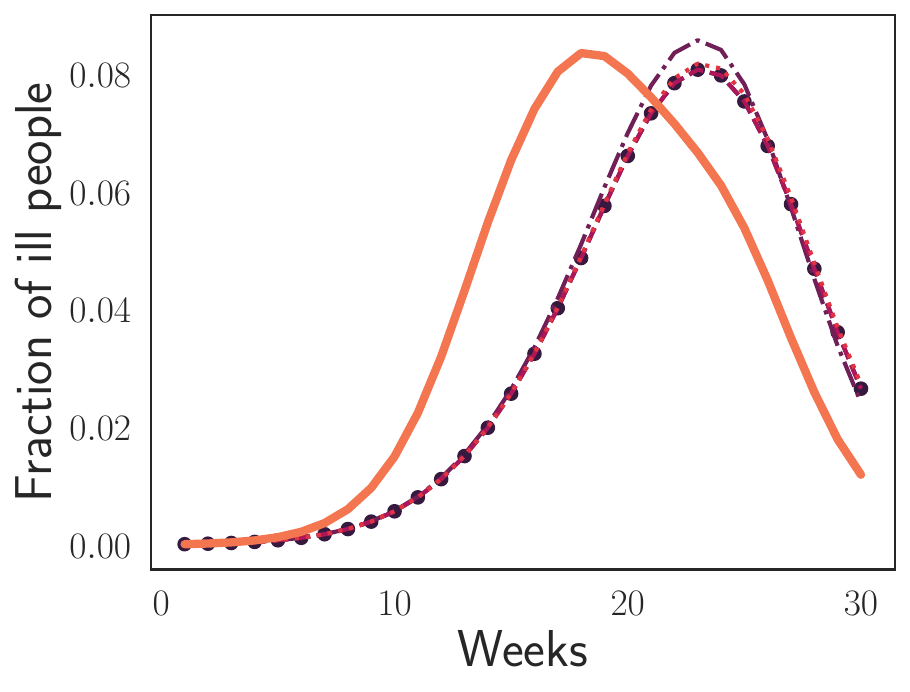}}\subfloat[Provinces, 2020/10/29]{\includegraphics[width=.5\columnwidth]{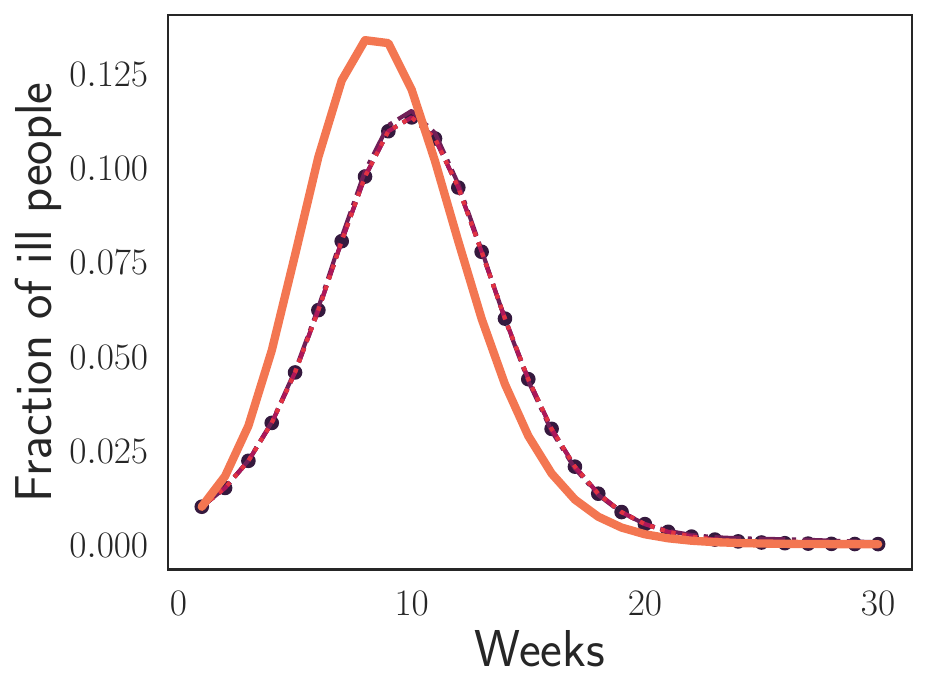}}\subfloat[Provinces, 2021/12/20]{\includegraphics[width=.5\columnwidth]{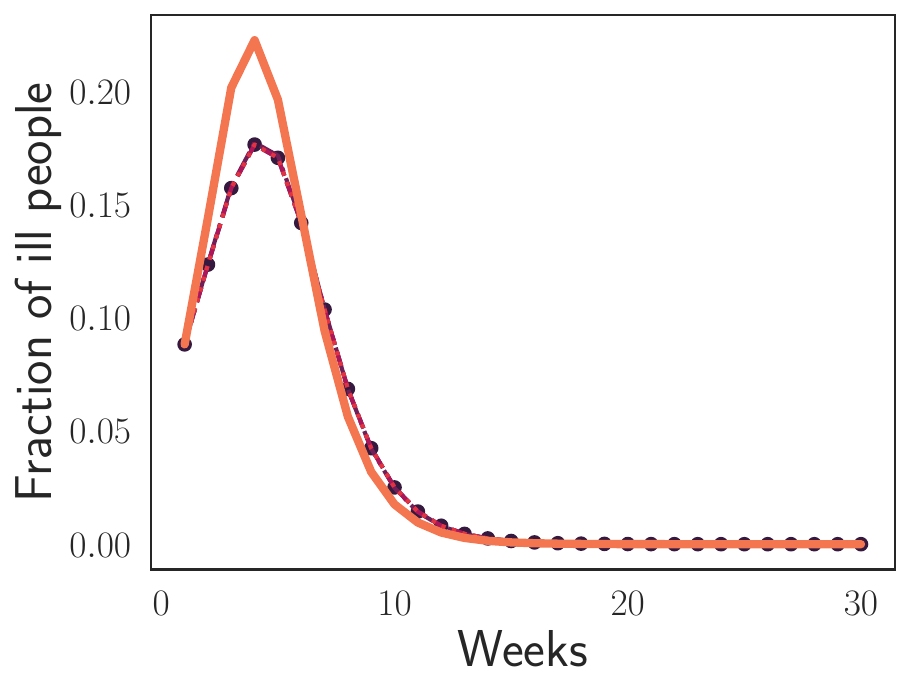}}
    \caption{Results of our numerical simulations (SIR). In (a--h), we report the plots obtained at the granularity of regions and provinces, respectively, with different starting dates. }
    \label{fig:sim2}
\end{figure*}

\begin{table*}
    \centering
    \resizebox{0.7\textwidth}{!}{
        \begin{tabular}{|c|c|cc|cc|cc|cc|}
           \hline
            \multirow{2}{*}{\textbf{Data}} & \multirow{2}{*}{\textbf{n}} & \multicolumn{2}{c|}{\textbf{SA}} & \multicolumn{2}{c|}{\textbf{TS}} & \multicolumn{2}{c|}{\textbf{GA}} &  \multicolumn{2}{c|}{\textbf{QA}} \\%& \multicolumn{2}{c|}{\textbf{Free}} \\
            & &  \textbf{p [\%]} & \textbf{a [\%]} &\textbf{p [\%]} & \textbf{a [\%]} & \textbf{p [\%]} & \textbf{a [\%]} & \textbf{p [\%]} & \textbf{a [\%]}%& \textbf{rate} & \textbf{area} 
            \\
            \hline
            2020/03/08 & 21 &  12.40 & 3.48 &  12.40 & 3.48 &   12.40 &  3.48 &  12.40 & 2.56 %& 0.13 & 1.20 
            \\
            2020/03/10 & 21 & 9.10 & 2.63 & 9.10 &  2.63 & 9.10 &  2.63 & 9.10 & 2.20 %&  0.13 & 1.20
            \\
            2020/10/29 & 21 &  12.77 & 1.93 & 12.77 & 1.93 &   12.77 & 1.93&   12.77 &  3.00 %& 0.14 & 1.20 
            \\
            2021/12/20 & 21 & 13.73 & 1.72 &   13.73 & 1.72 &   13.73 & 1.72 &   13.73 & 10.05 %& 0.16 & 1.20 
            \\
\hline
            2020/03/08 & 107 &  4.93 & 17.96 &   4.93 &
 17.96  &  -1.48 & 14.31 &  3.61 & 16.66 %& 0.08 & 1.10 
            \\
            2020/03/10 & 107 &  3.37 &  14.79 &   3.37 &  14.79  & -2.61 & 11.99 & 2.15 & 13.82 %& %0.08 & 1.12
            \\
            2020/10/29 & 107 &  15.31 &  9.02 &  15.31 &  9.02 &  14.02 & 7.74 &  15.29 &  9.02 %& 0.13 & 1.16 
            \\
            2021/12/20 & 107 &  20.70 &  7.54 &  20.70 &  7.54 & 20.31 & 7.34 &  20.70 &   7.54 %& 0.22 & 1.18 
            \\
            \hline
        \end{tabular}
    }
    \caption{Performance obtained with different methods in terms of reduction of the peak (p) and average infected/day (a) with respect to the uncontrolled dynamics, considering SIR model.}
    \label{tab:SIRPicco}
\end{table*}

\begin{table*}[h]
    \centering
    \resizebox{0.8\textwidth}{!}{ 
         \begin{tabular}{|c|c|cc|cc|cc|cc|}
             \hline
             \multirow{2}{*}{\textbf{Data}} & \multirow{2}{*}{\textbf{N}} & \multicolumn{2}{c|}{\textbf{SA}} & \multicolumn{2}{c|}{\textbf{TS}} & \multicolumn{2}{c|}{\textbf{GA}} &  \multicolumn{2}{c|}{\textbf{QA}} \\
            & & \textbf{t [\si{\milli\second}]} & \textbf{C} & \textbf{t [\si{\milli\second}]} & \textbf{C} & \textbf{t [\si{\milli\second}]} & \textbf{C} & \textbf{t [\si{\milli \second}]} & \textbf{C}\\
             \hline
             2020/03/08 & 21 & 2.07 & 61.71e6 & 2112.88 & 61.71e6 & 55.92 & 61.71e6 & 0.04 & 61.81e6 \\
             2020/03/10 & 21 & 1.94 & 61.91e6 & 2103.59 &  61.91e6 & 49.50 &   61.95e6 & 0.04 &  62.13e6 \\
             2020/10/29 & 21 & 1.93 & 51.10e6 & 2110.27 & 51.10e6 & 51.32 &  51.10e6 & 0.04 & 51.24e6 \\
             2021/12/20 & 21 & 1.89 & 39.99e6 & 2119.29 & 39.99e6 & 68.24 & 39.99e6 & 0.04 & 40.50e6 \\   
\hline
             2020/03/08 & 107 & 784.40 & 61.39e6 & 2124.64 & 61.39e6 & 508.54 & 47.05e6 & 0.04 & 56.96e6 \\
             2020/03/10 & 107 & 1072.64 & 61.89e6 & 2126.41 & 61.89e6 & 440.78 &  47.09e6 & 0.04 & 57.38e6 \\
             2020/10/29 & 107 & 824.93 & 59.42e6 & 2124.39  & 59.42e6 & 495.81 &  47.27e6 & 0.04 & 58.74e6 \\
             2021/12/20 & 107 & 1306.50 & 43.38e6 & 2116.03 & 43.38e6 & 371.79 & 36.23e6 & 0.04 & 43.37e6 \\     
             \hline
         \end{tabular}}
    \caption{Average solving time for the different methods, considering SIR model.}
    \label{tab:SIR}
\end{table*}

\begin{figure*}
    \centering
\subfloat[North, 2020/03/08]{\includegraphics[width=.5\columnwidth]{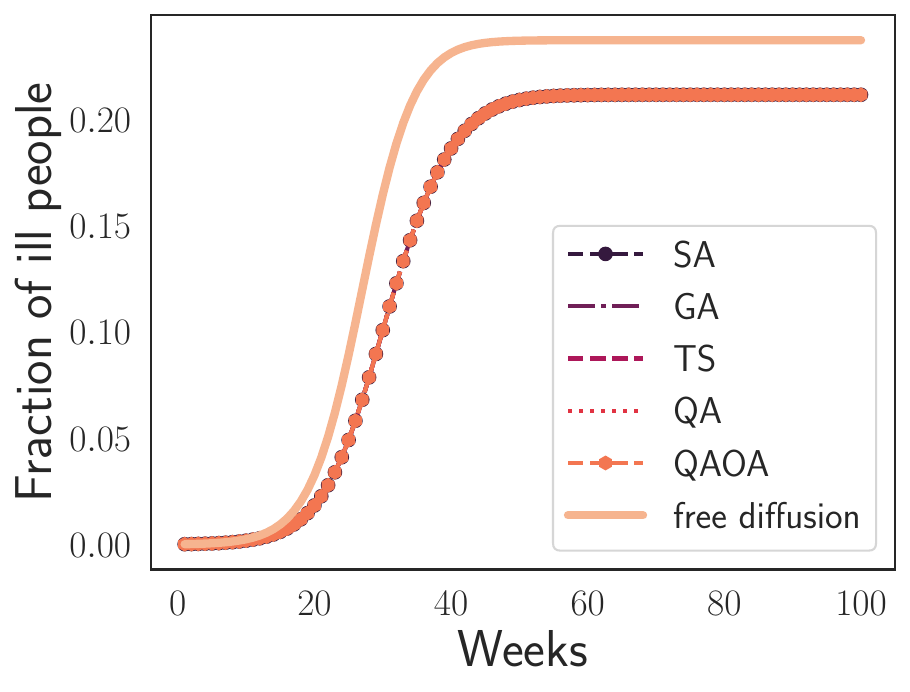}}
\subfloat[North, 2020/03/10]{\includegraphics[width=.5\columnwidth]{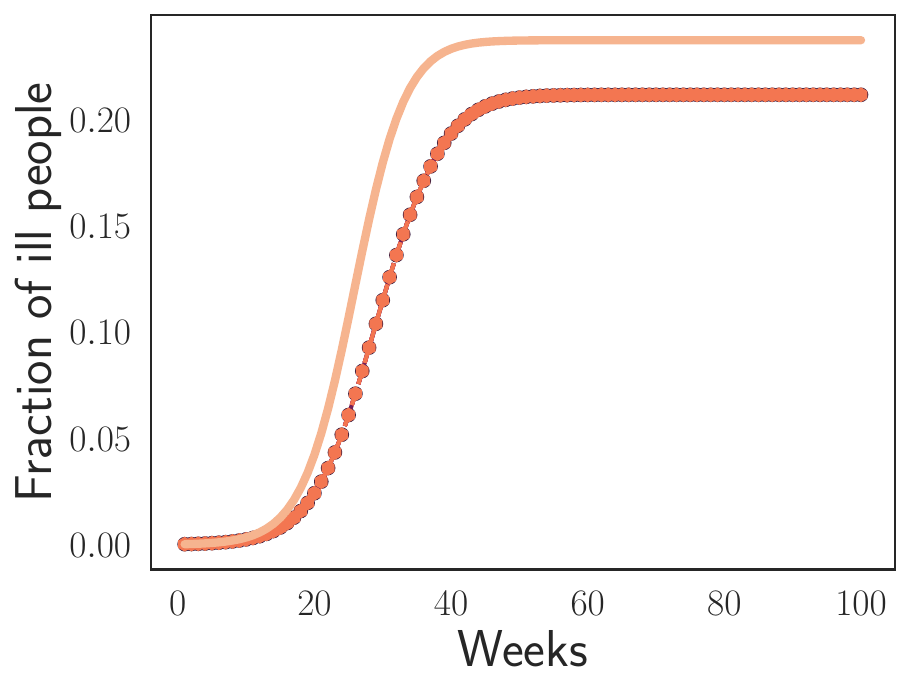}}\subfloat[North, 2020/10/29]{\includegraphics[width=.5\columnwidth]{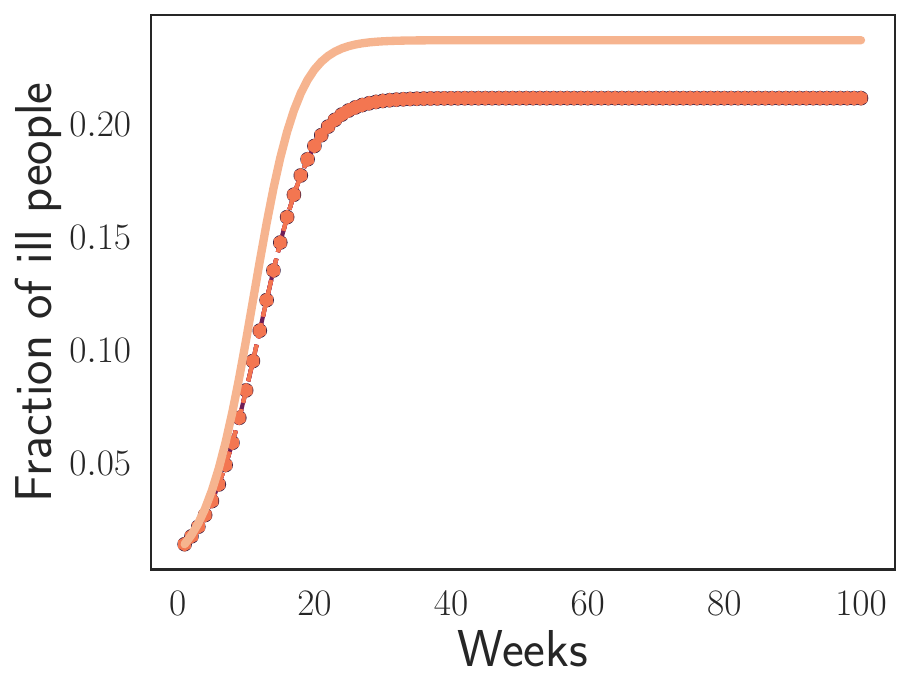}}\subfloat[North, 2021/12/20]{\includegraphics[width=.5\columnwidth]{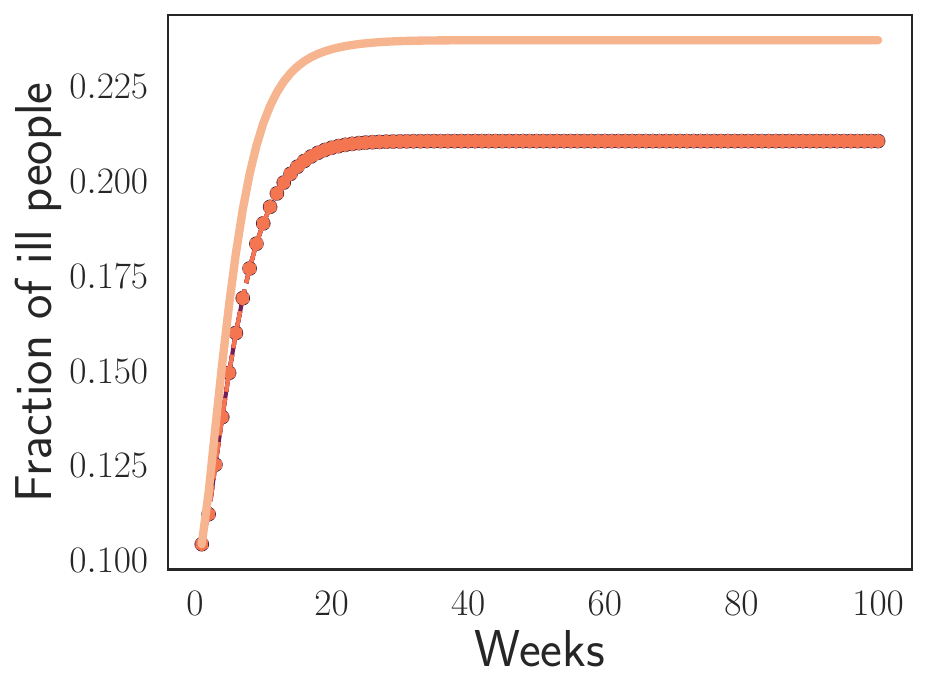}}\\[1ex]\subfloat[Center, 2020/03/08]{\includegraphics[width=.5\columnwidth]{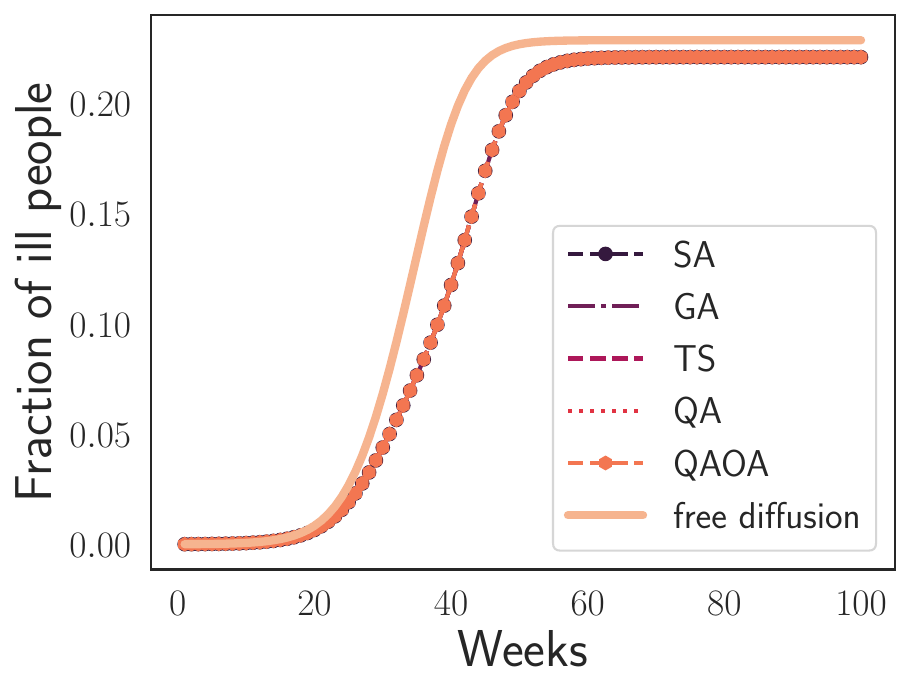}}
\subfloat[Center, 2020/03/10]{\includegraphics[width=.5\columnwidth]{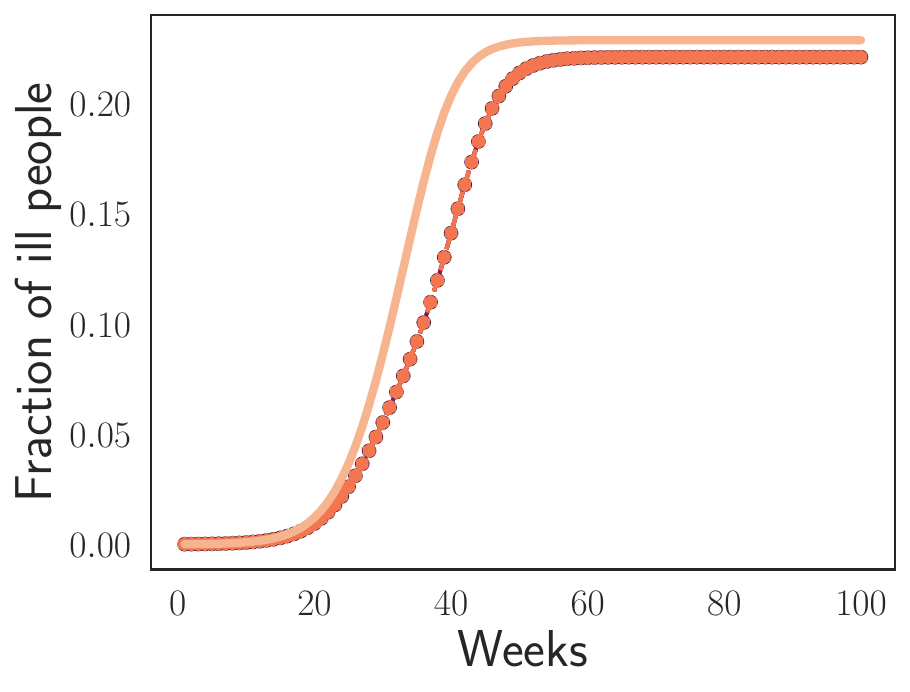}}\subfloat[Center, 2020/10/29]{\includegraphics[width=.5\columnwidth]{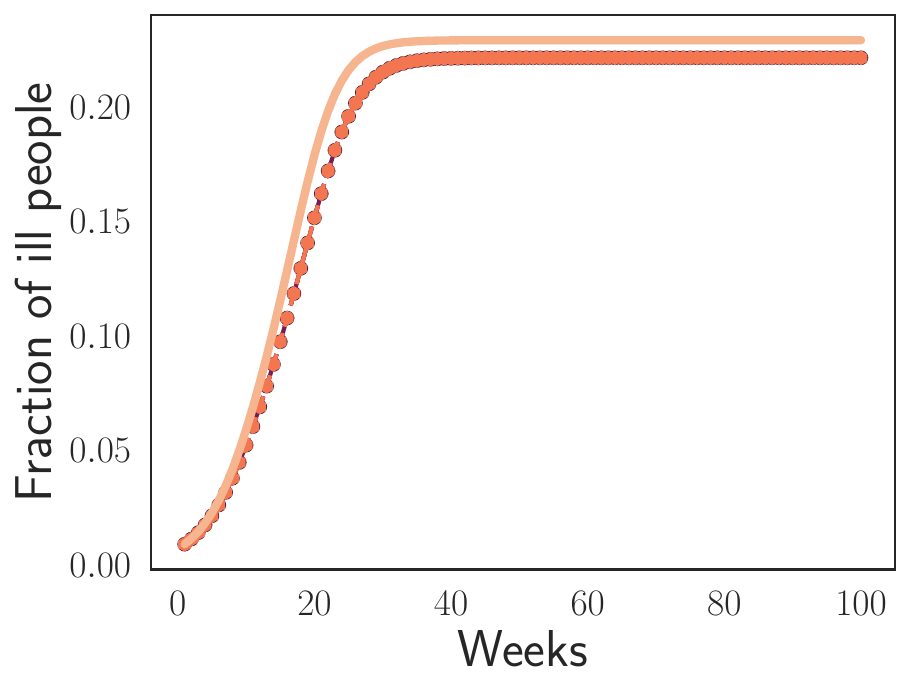}}\subfloat[Center, 2021/12/20]{\includegraphics[width=.5\columnwidth]{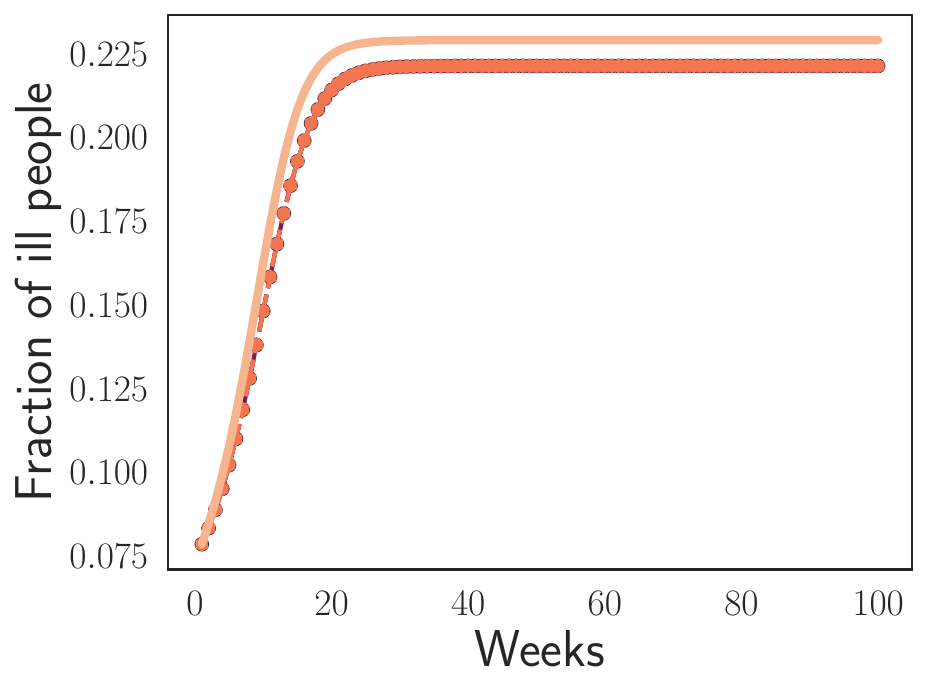}}
\\[1ex]\subfloat[South, 2020/03/08]{\includegraphics[width=.5\columnwidth]{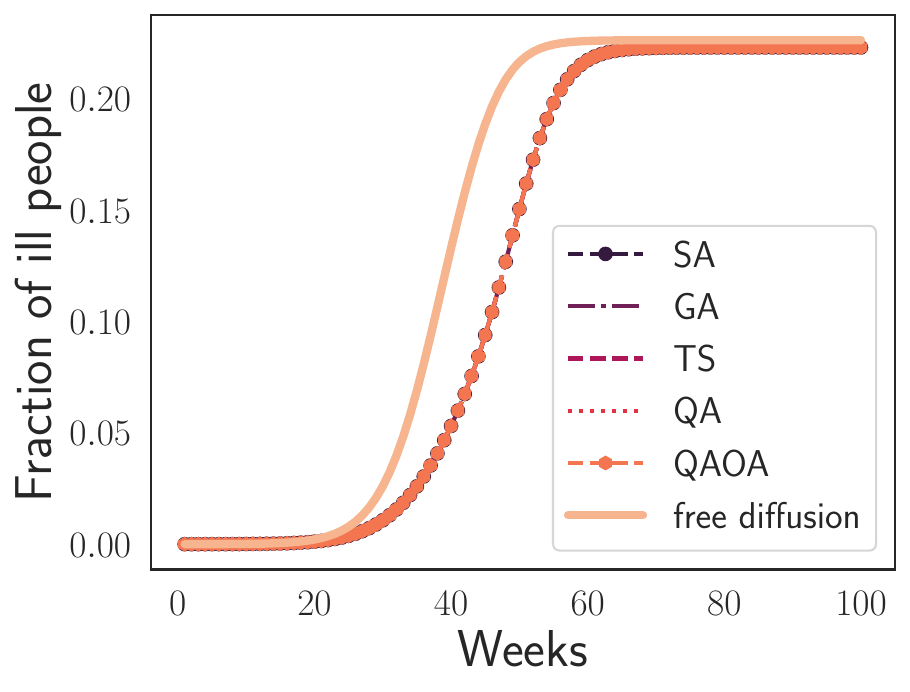}}
\subfloat[South, 2020/03/10]{\includegraphics[width=.5\columnwidth]{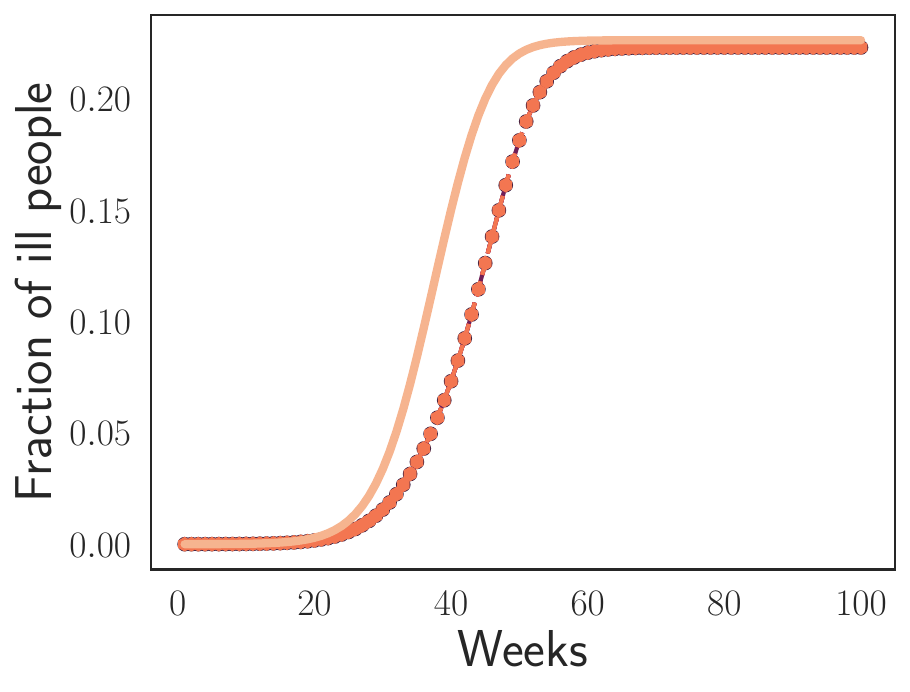}}\subfloat[South, 2020/10/29]{\includegraphics[width=.5\columnwidth]{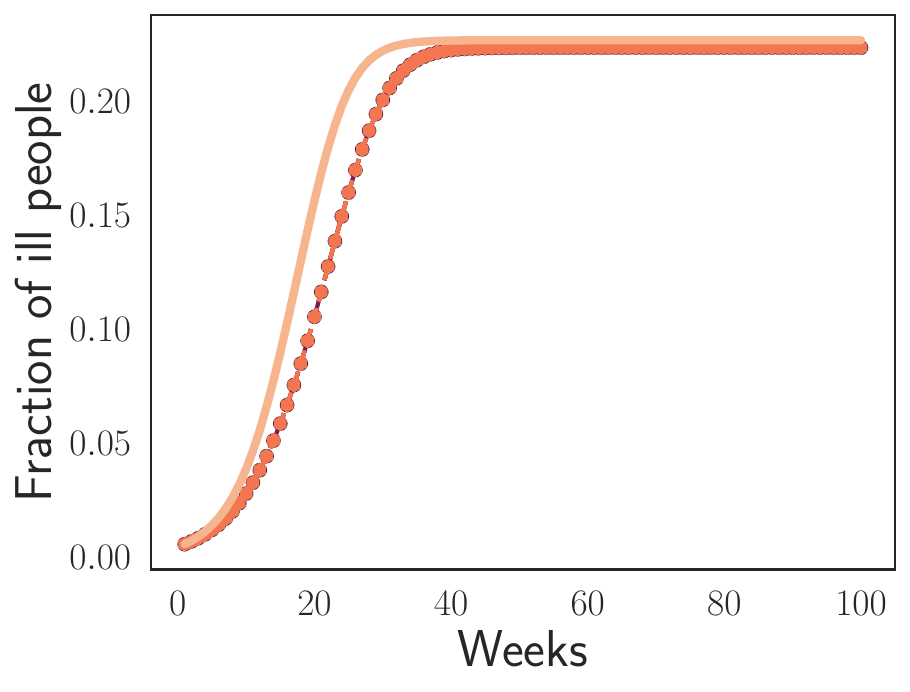}}\subfloat[South, 2021/12/20]{\includegraphics[width=.5\columnwidth]{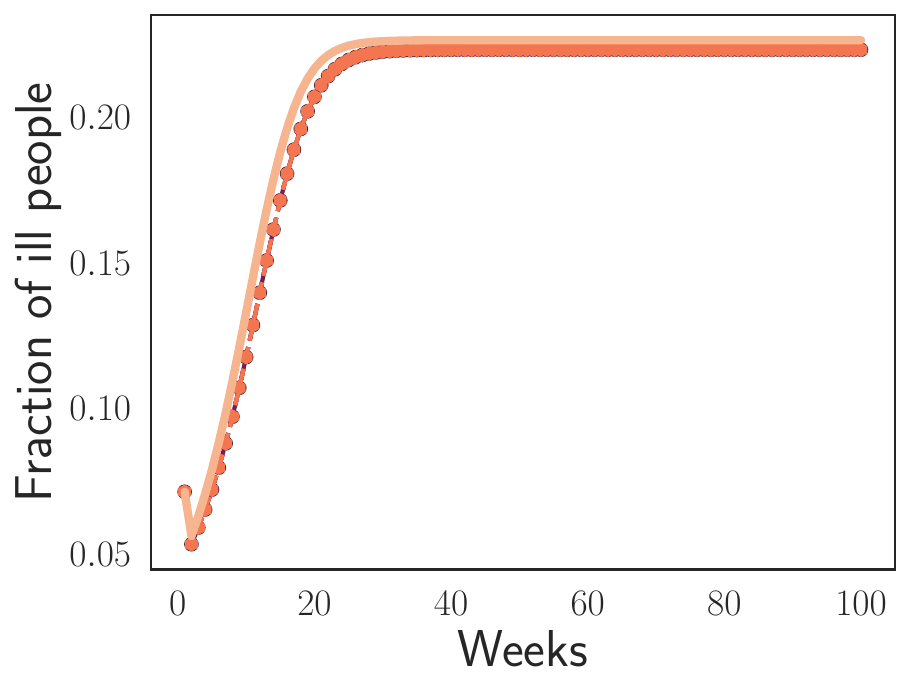}}
    \caption{Results of our numerical simulations. In (a--h), we report the plots obtained at the granularity of regions in the three parts of Italy (North, Center, South) with different starting dates, considering SIS. }
    \label{fig:sim4}
\end{figure*}

Figure~\ref{fig:sim4} presents the results obtained at the regional level for the three main areas of Italy—North, Center, and South—using the SIS epidemic model. This experiment demonstrates the feasibility of solving the control problem using the QAOA algorithm, which achieves solution quality comparable to that of the other solvers. Furthermore, the results across both regional and provincial levels confirm that the higher the granularity and dimensionality of the network, the greater the positive impact of the proposed control approach.

Similarly, the results for the SIR epidemic diffusion model at both the regional and provincial levels are presented in Figure~\ref{fig:sim2}. These plots again prove the effectiveness of all control methods in reducing the number of infected individuals compared to the uncontrolled baseline. It is possible to notice that the effectiveness of the control mechanism is strongly influenced by the initial conditions. For instance, in the scenarios presented in (e) and (f), the control primarily delayed the evolution of the pandemic, with negligible impact on peak reduction.

Quantitative comparisons are reported in Table~\ref{tab:SIRPicco}, showing that QA maintains a performance level comparable to classical approaches in most settings. These outcomes are also achieved with significantly reduced computational time, as shown in Table~\ref{tab:SIR}. As in the SIS case, QA consistently delivers solutions much faster than classical methods, particularly in large-scale scenarios, without compromising the quality of the results. This highlights QA’s potential as a valuable tool for timely intervention planning during epidemic outbreaks.

\begin{figure*}
    \centering
\subfloat[North, 2020/03/08]{\includegraphics[width=.5\columnwidth]{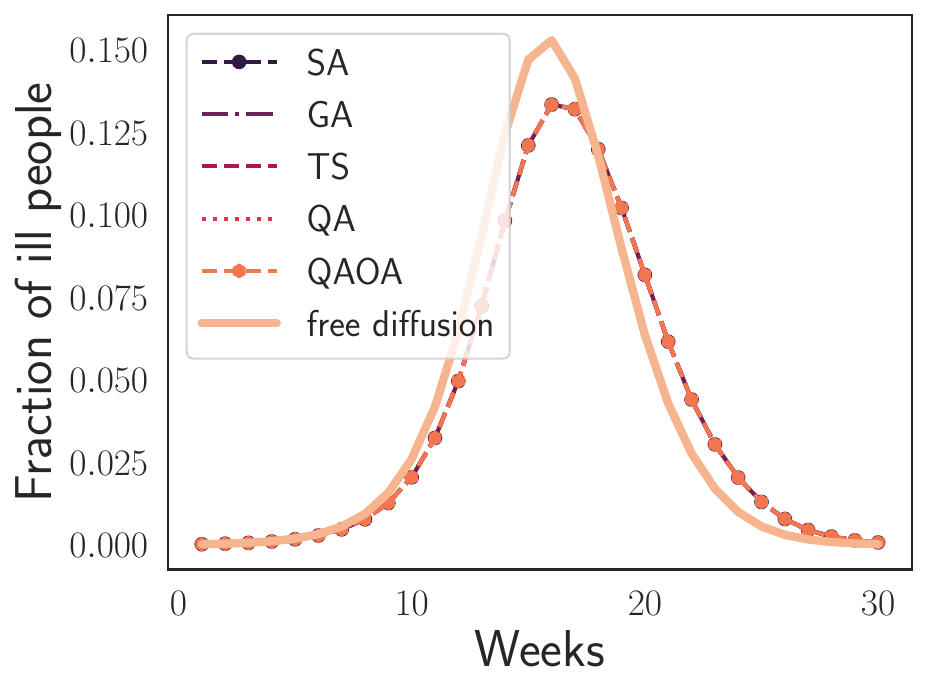}}
\subfloat[North, 2020/03/10]{\includegraphics[width=.5\columnwidth]{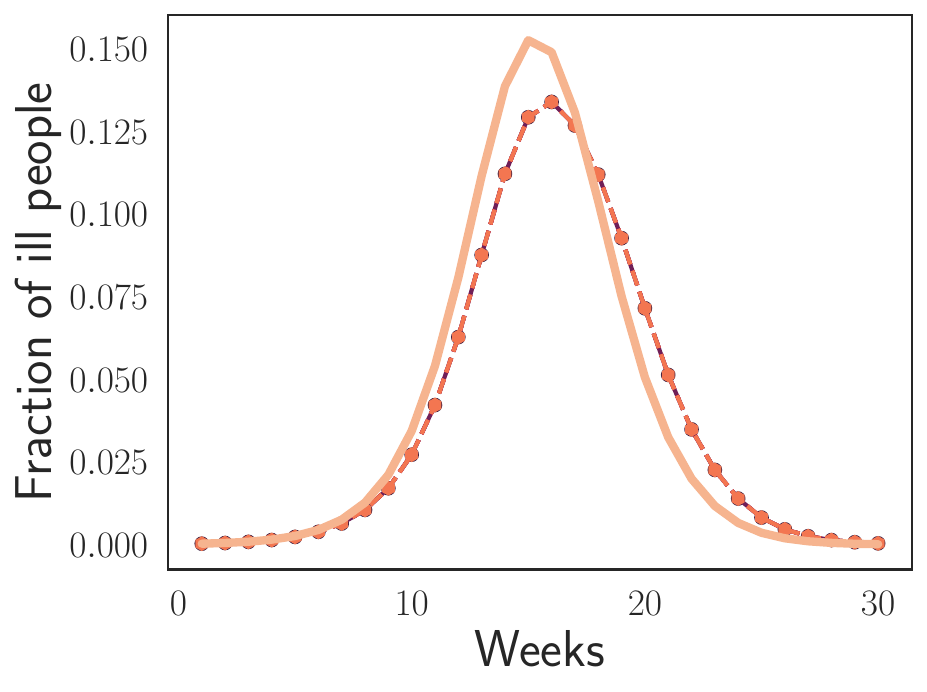}}\subfloat[North, 2020/10/29]{\includegraphics[width=.5\columnwidth]{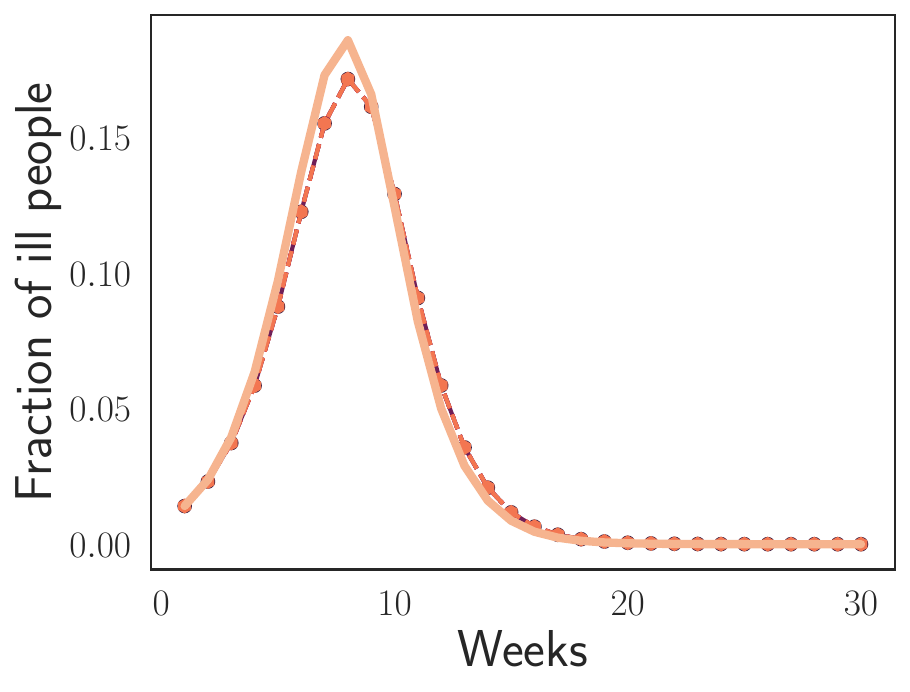}}\subfloat[North, 2021/12/20]{\includegraphics[width=.5\columnwidth]{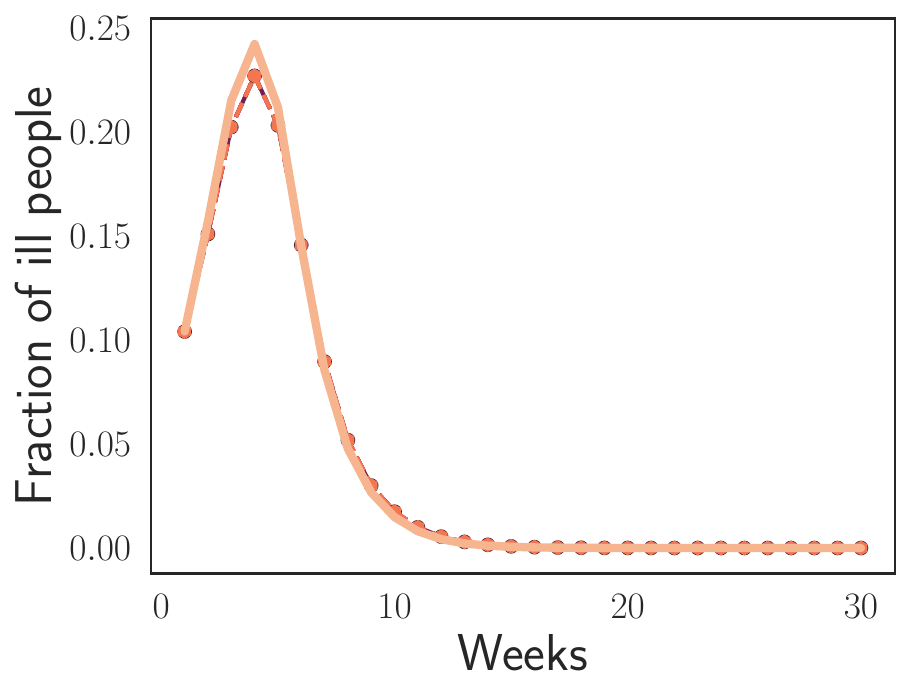}}\\[1ex]\subfloat[Center, 2020/03/08]{\includegraphics[width=.5\columnwidth]{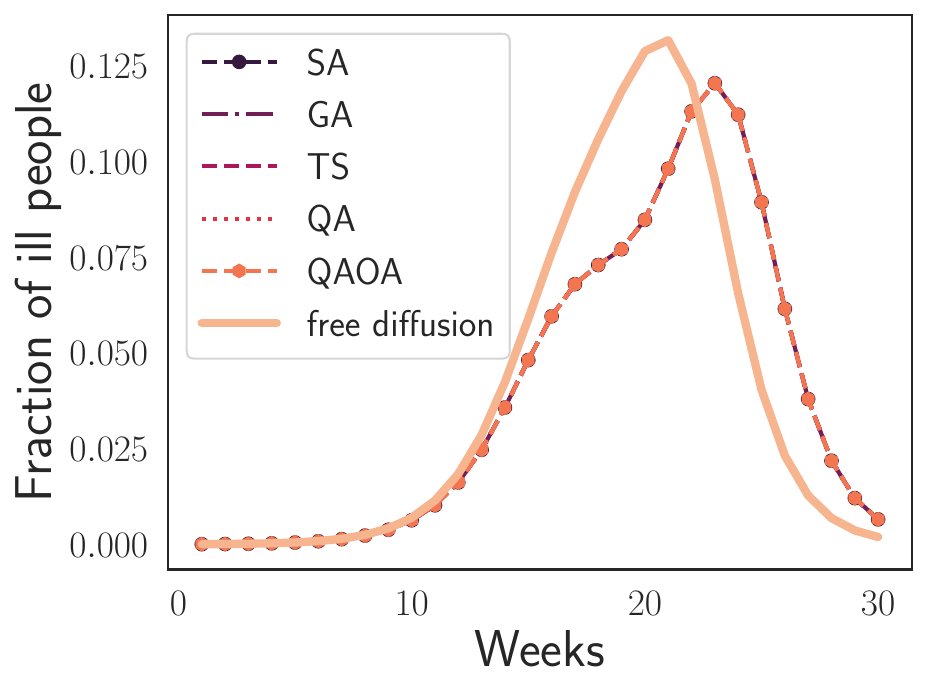}}
\subfloat[Center, 2020/03/10]{\includegraphics[width=.5\columnwidth]{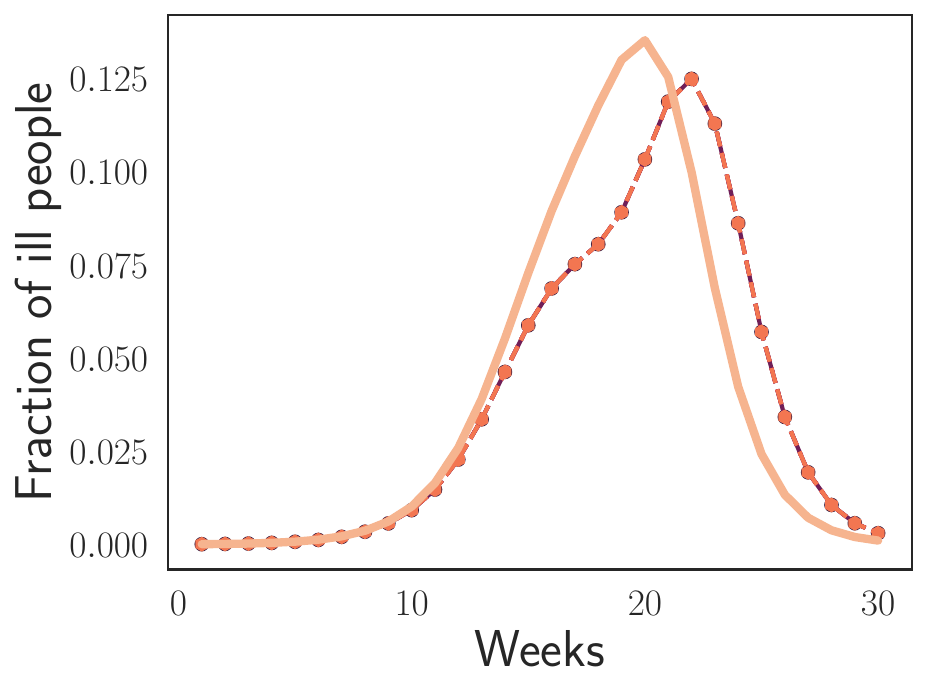}}\subfloat[Center, 2020/10/29]{\includegraphics[width=.5\columnwidth]{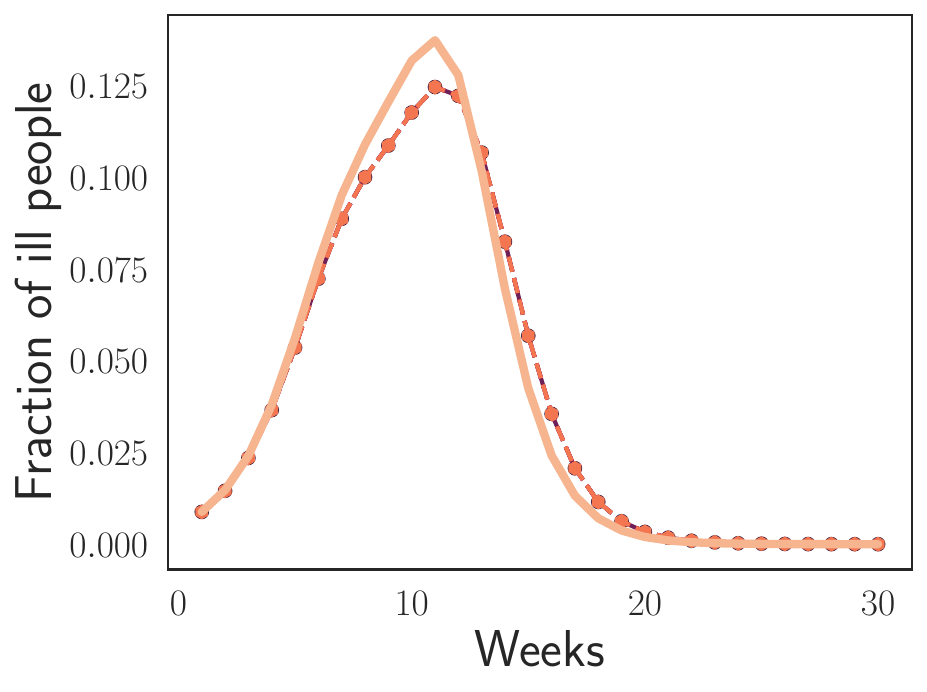}}\subfloat[Center, 2021/12/20]{\includegraphics[width=.5\columnwidth]{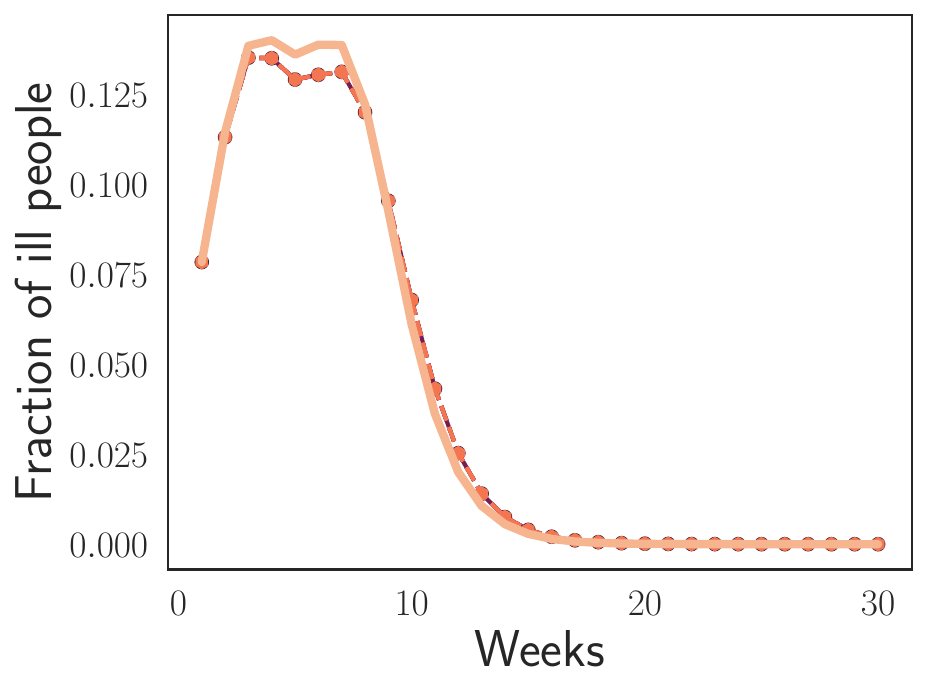}}
\\[1ex]\subfloat[South, 2020/03/08]{\includegraphics[width=.5\columnwidth]{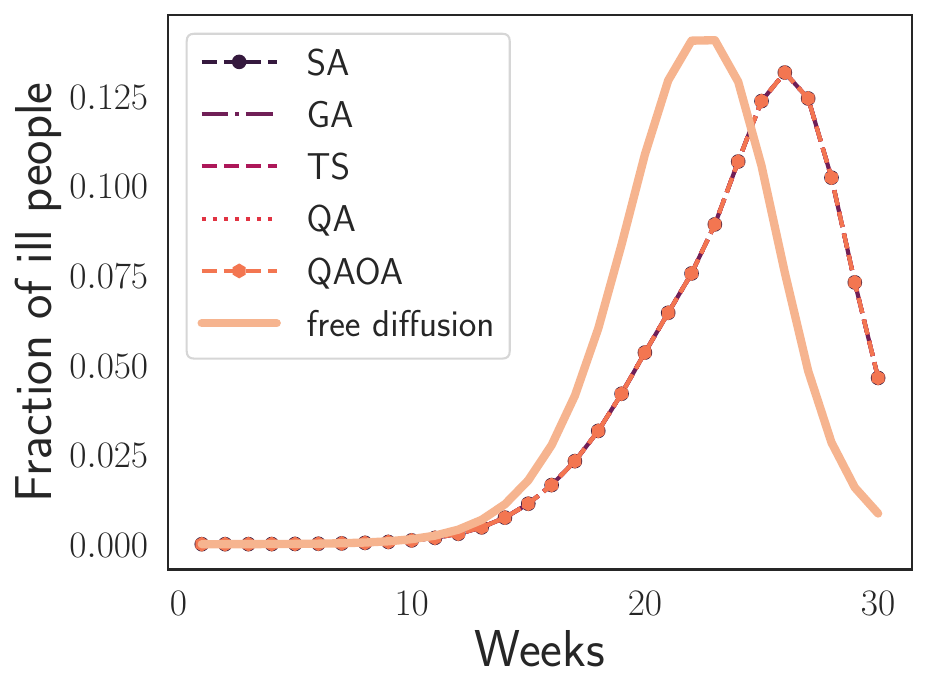}}
\subfloat[South, 2020/03/10]{\includegraphics[width=.5\columnwidth]{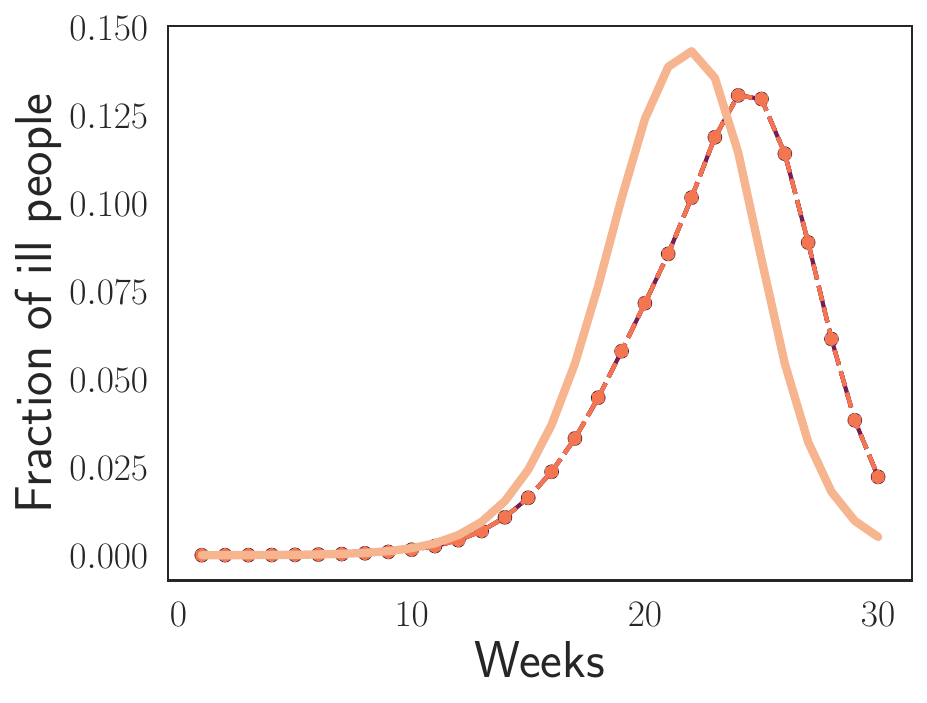}}\subfloat[South, 2020/10/29]{\includegraphics[width=.5\columnwidth]{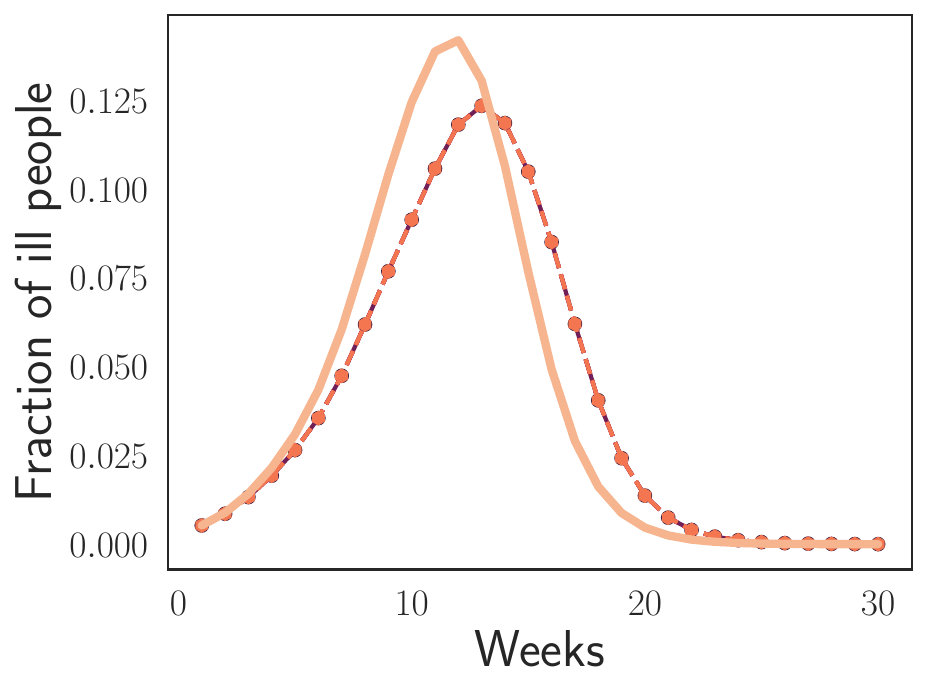}}\subfloat[South, 2021/12/20]{\includegraphics[width=.5\columnwidth]{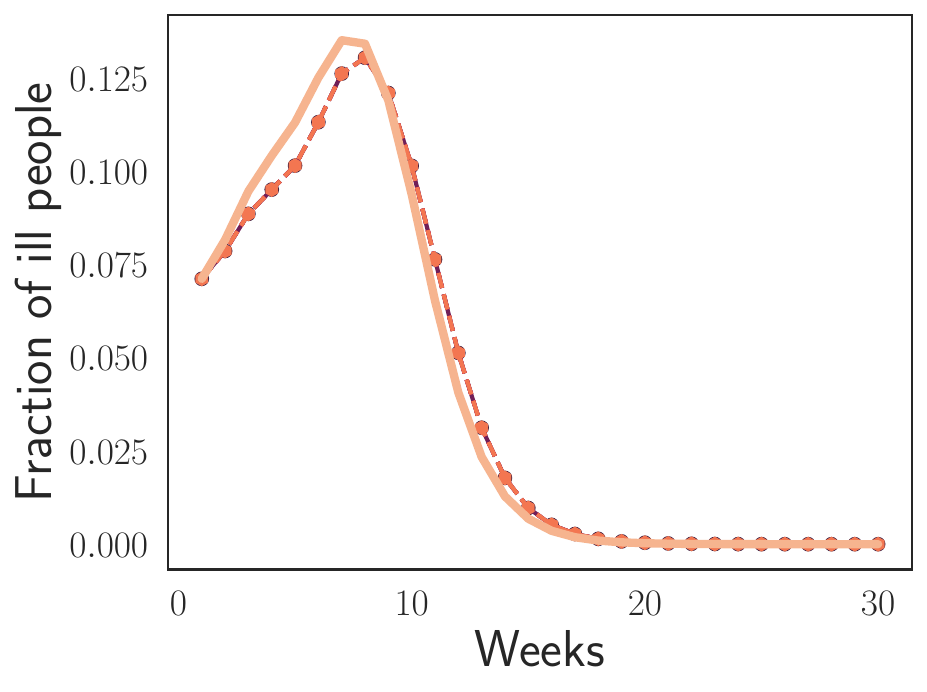}}
    \caption{Results of our numerical simulations. In (a--h), we report the plots obtained at the granularity of regions in the three parts of Italy (North, Center, South) with different starting dates, considering SIR. }
    \label{fig:sim5}
\end{figure*}

Furthermore, Figure~\ref{fig:sim5} illustrate the results obtained at the regional level for the three main areas of Italy --- North, Center, and South --- considering the SIR epidemic model. This test proves the feasibility of solving the control problem using the QAOA algorithm, which reaches solution quality comparable to that of the other solvers. Also in this case, the results across both regional and provincial levels confirm that the higher the granularity and dimensionality of the network, the greater the positive impact of the proposed control approach.

\begin{figure*}
    \centering
\subfloat[SIR, 2020/03/08]{\includegraphics[width=.5\columnwidth]{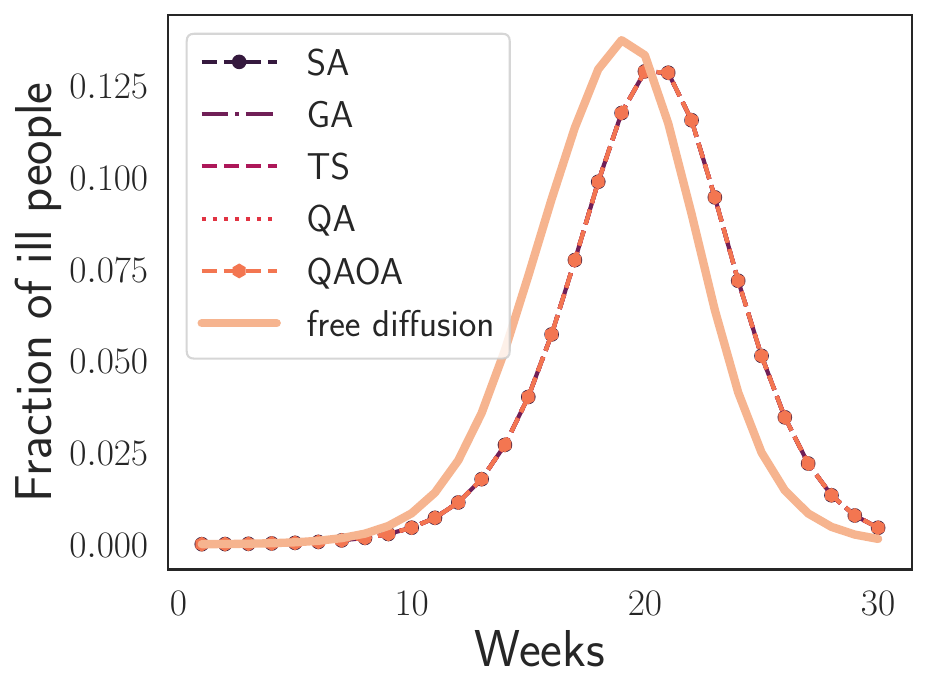}}
\subfloat[SIR, 2020/03/10]{\includegraphics[width=.5\columnwidth]{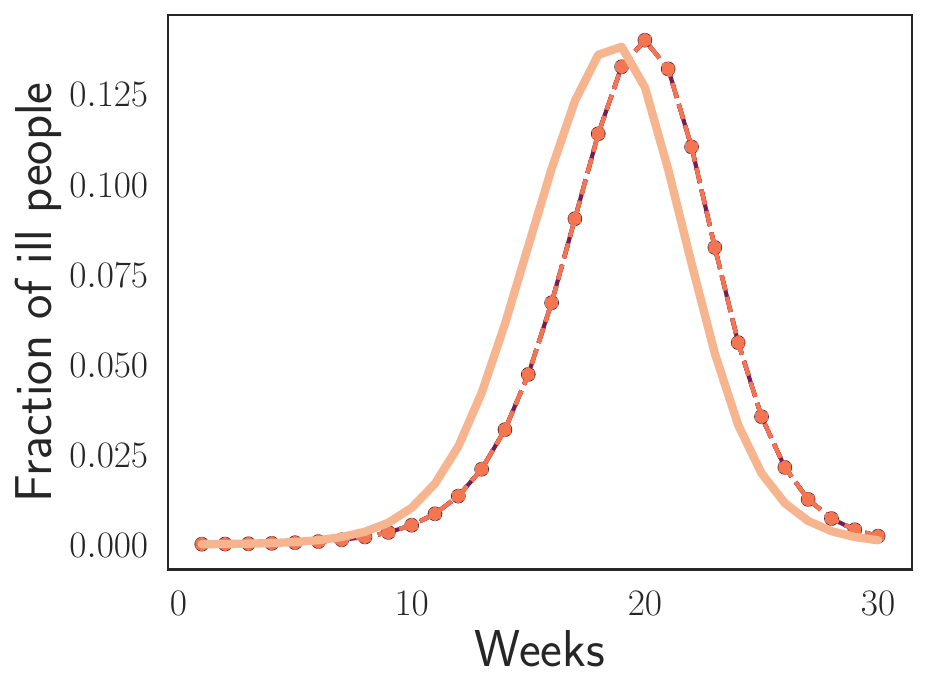}}\subfloat[SIR, 2020/10/29]{\includegraphics[width=.5\columnwidth]{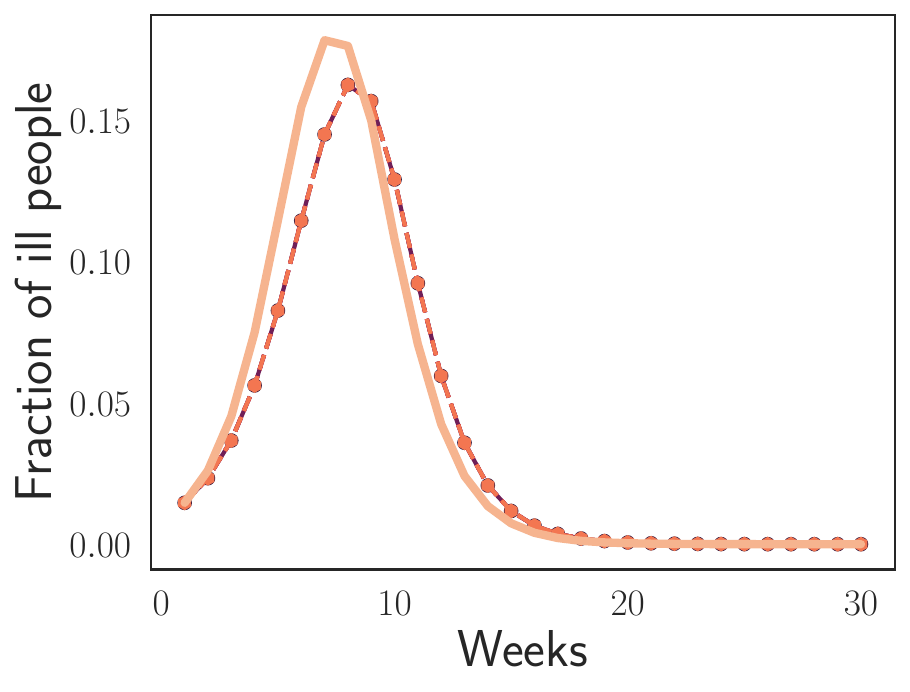}}\subfloat[SIR, 2021/12/20]{\includegraphics[width=.5\columnwidth]{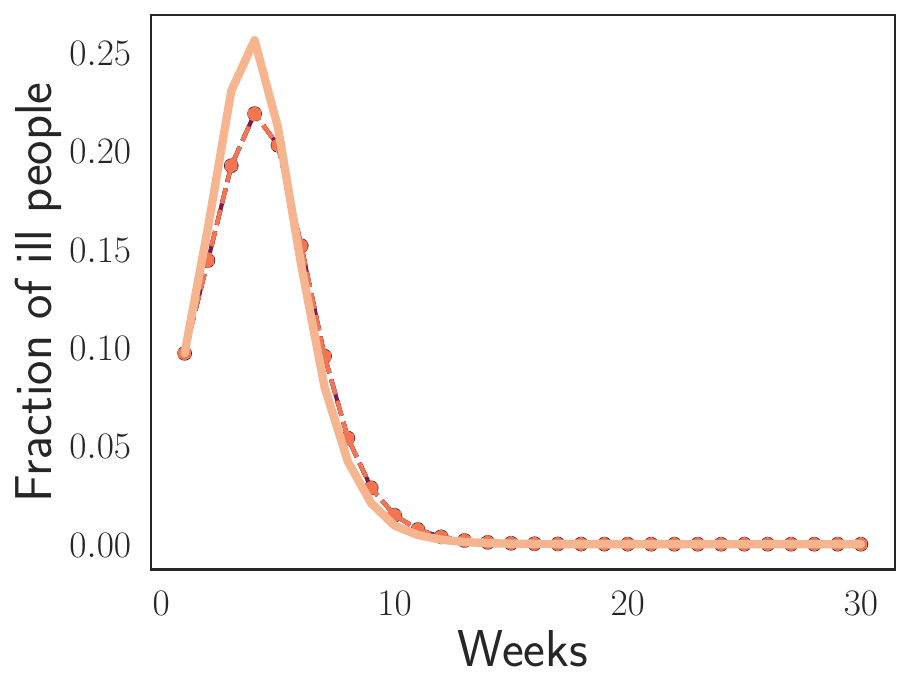}}\\[1ex]\subfloat[SIS, 2020/03/08]{\includegraphics[width=.5\columnwidth]{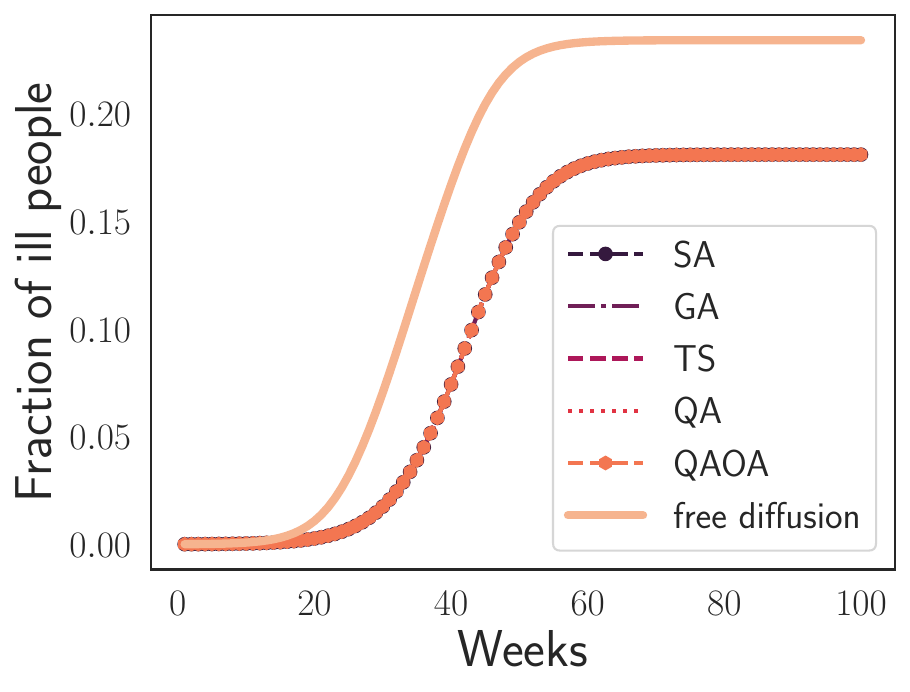}}
\subfloat[SIS, 2020/03/10]{\includegraphics[width=.5\columnwidth]{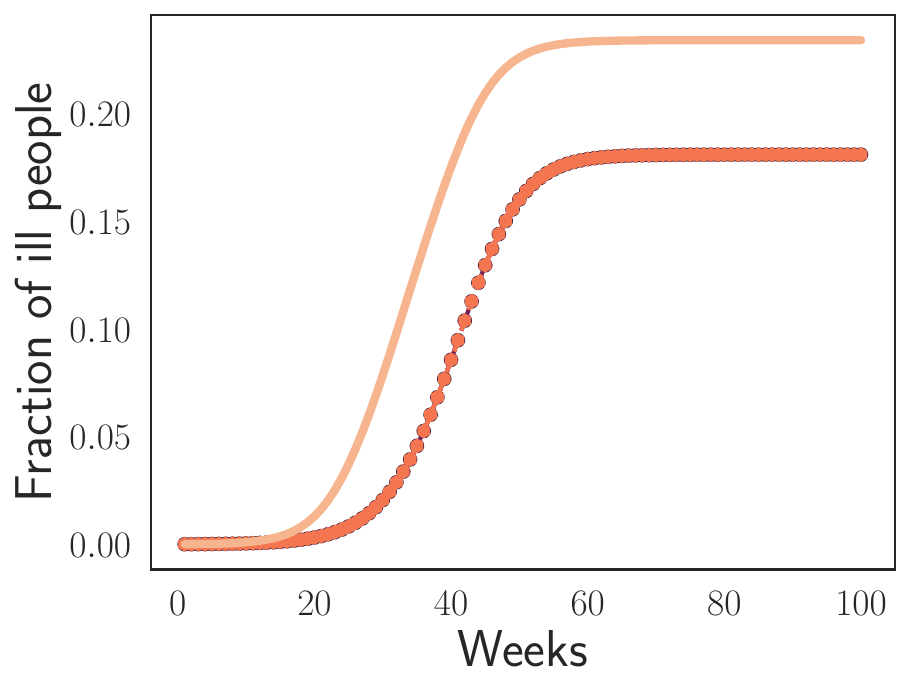}}\subfloat[SIS, 2020/10/29]{\includegraphics[width=.5\columnwidth]{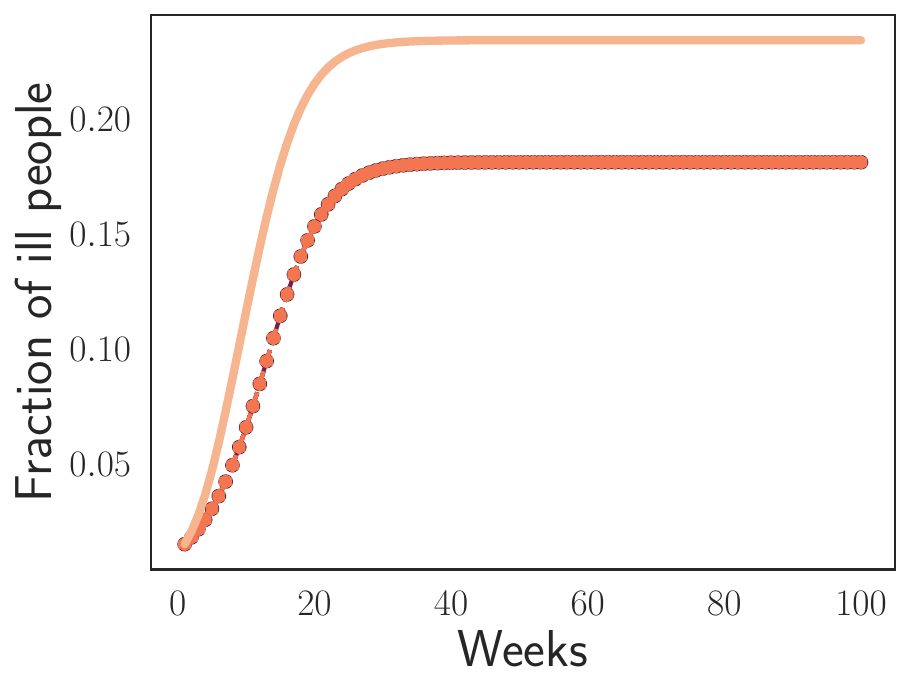}}\subfloat[SIS, 2021/12/20]{\includegraphics[width=.5\columnwidth]{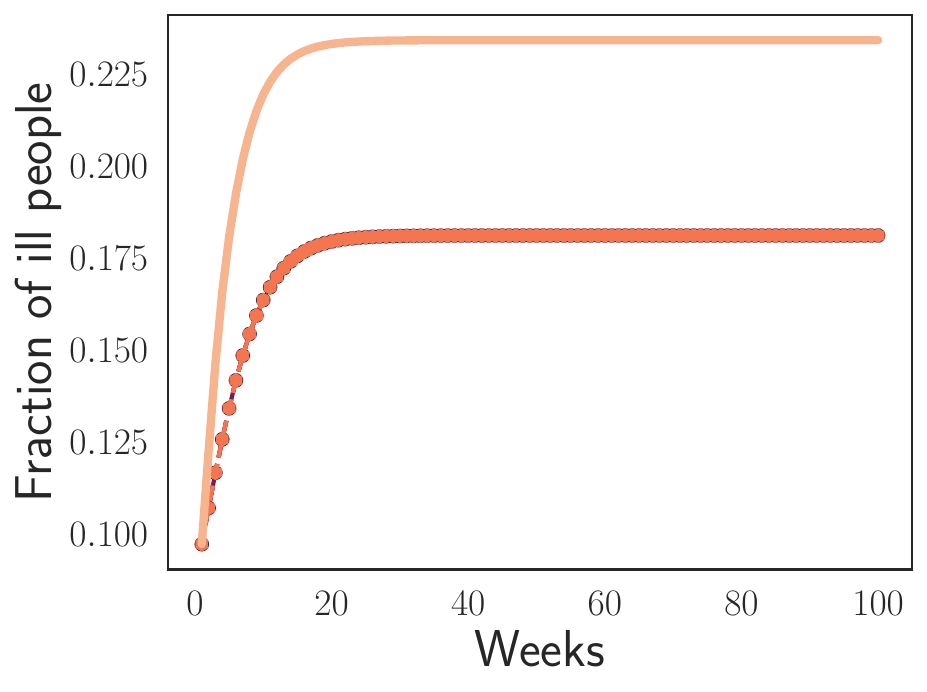}}
    \caption{Results of our numerical simulations. In (a--d) and (e--h), we report the plots obtained at the granularity of single region province (Piedmont) with different starting dates, considering SIS and SIR models. }
    \label{fig:sim6}
\end{figure*}

Finally, Figure \ref{fig:sim6} illustrates the results obtained at the provincial level considering a single region (Piedmont) for both the SIS and the SIR evolution mechanism. The benefits of the QUBO-based control mechanism in this context are more pronounced in the SIS model. In the absence of an immunization mechanism, mobility restrictions play a crucial role in preventing the system from stabilizing at a high level of infection. This increases the impact of optimal control strategies, making the efficiency of QA particularly advantageous.

\subsection{Computation Time and Scalability}

\begin{figure}
    \centering
\includegraphics[width=\columnwidth]{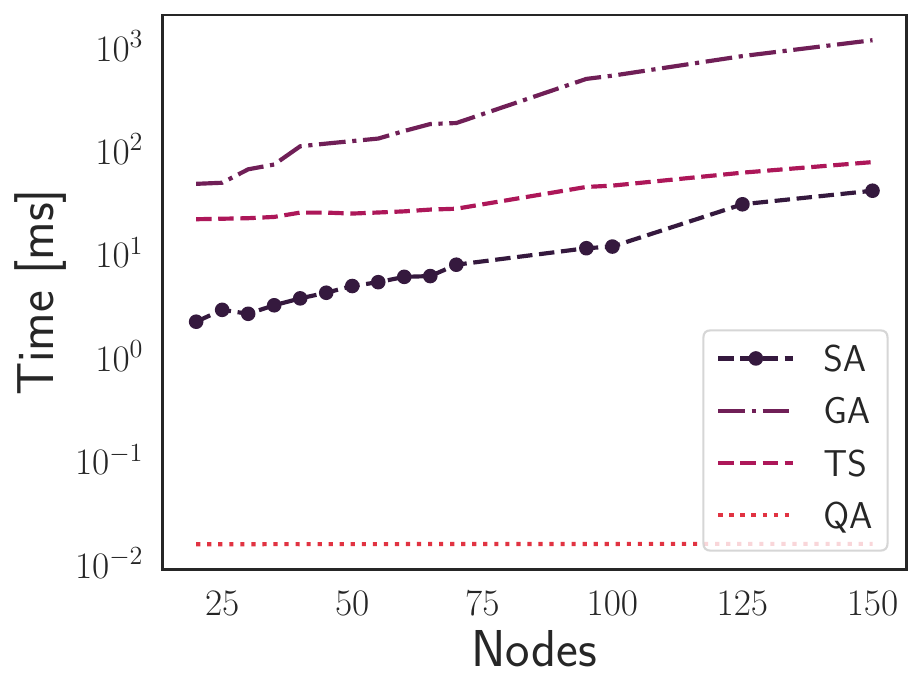}
    \caption{Solving time of the solvers as a function of the number of nodes, considering the SIR model with synthetic data. }
    \label{fig:Time}
\end{figure}

\begin{figure}
    \centering
\includegraphics[width=\columnwidth]{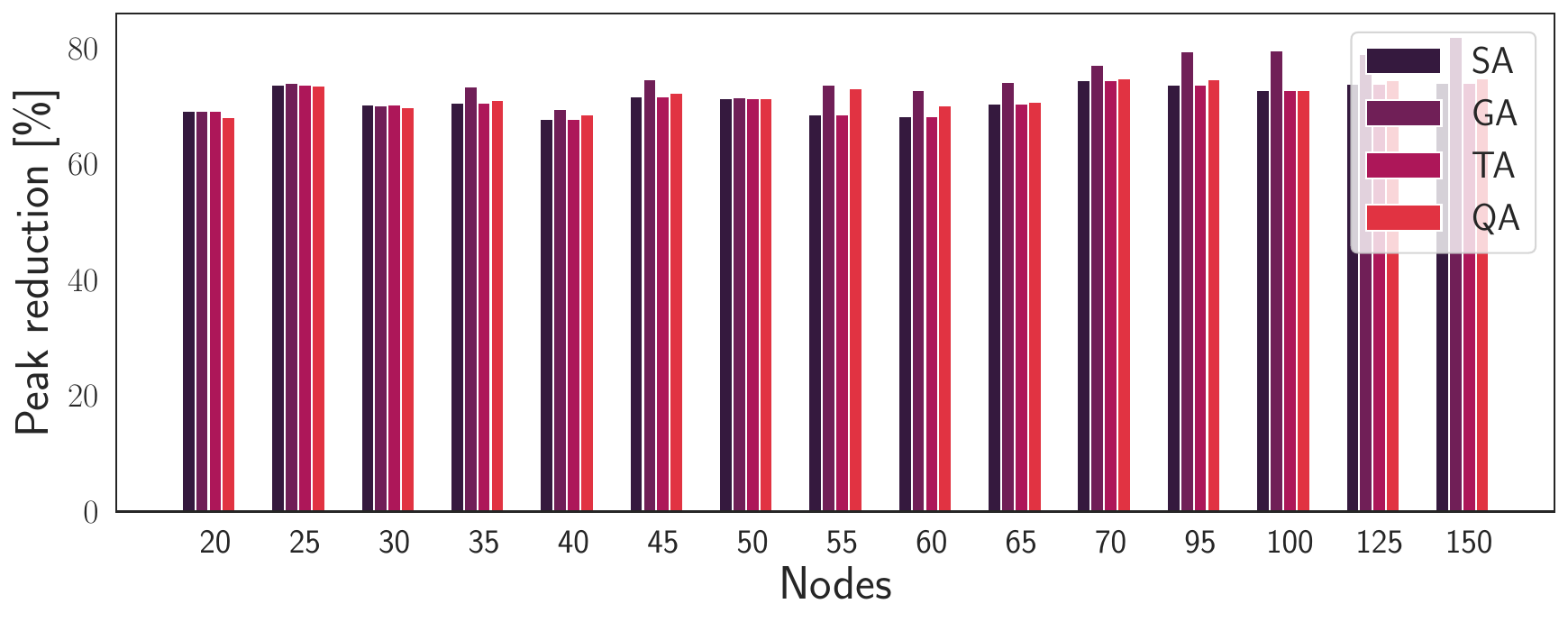}
    \caption{Average peak (p) reduction of the solvers as a function of the number of nodes, considering the SIR model with synthetic data. }
    \label{fig:peak}
\end{figure}

The solving times attained using the synthetic data are reported in Figure~\ref{fig:Time} as a function of the network size, which spans from $10$ to $150$ nodes. The network and its corresponding adjacency matrix, used to model population mobility across nodes, are generated randomly to have a connection density of $30\%$, i.e. each pair of nodes has a $30\%$ probability of being connected. The initial infection rate is assigned a random value between $0.01\%$ and $0.1\%$ of the population of each node, in accordance with the COVID-19 real data. The solution quality is assessed in terms of peak reduction ($p$), as shown in Figure~\ref{fig:peak}. QA exhibits an almost constant solving time ($\sim\SI{0.04}{\milli\second}$), independent of the network scale, in contrast to classical solvers, whose solving times increase rapidly with the problem size. At the same time, the quality of the solutions provided by QA is very close to that of the classical solvers, demonstrating that it offers a good trade-off between computational efficiency and solution quality.

\section{Discussion}\label{sec:discussion}

The obtained results prove the advantages of the proposed QUBO formulation and of quantum solvers in epidemic network control. 

The compact encoding of the QUBO cost function allowed these solvers to explore optimal node-isolation strategies even in high-dimensional spaces. Our formulation supports both SIR and SIS dynamics, showcasing flexibility in modeling diseases with temporary or permanent immunity. This formulation offers a flexible foundation that can be readily extended in the future to support diverse and more refined models. The integration of economic weight ($\gamma$) into the cost function ensures that isolation is only applied when justified by epidemiological severity, enabling cost-aware strategies.

While all solvers reduced infection spread, QA  provided solutions that guarantee a good balance between containment effectiveness and computational efficiency. 
Specifically, QA required only milliseconds for problems where GA or TS took seconds, making it more suitable for rolling-horizon or adaptive epidemic policy planning. The results obtained via QAOA further confirm that gate-based quantum devices can support epidemic control problems. While still constrained by qubit count and noise, these results demonstrate the feasibility of circuit-based quantum decision-making.

Finally, our method adapts across administrative levels (regions, provinces) and maintains performance consistency. This suggests its practical applicability for multi-level governance in national and international contexts.

\section{Conclusions}\label{sec:conclusions}

In this paper, we introduced a quantum-based approach for optimal intervention in epidemic processes on networks. By formulating the transient epidemic control problem as a QUBO, we ensured compatibility with both quantum annealers and gate-based quantum algorithms. We implemented the control strategy using a rolling-horizon scheme for two different classical epidemic models (the SIS and the SIR models), and we benchmarked the performance of five solvers, including a quantum annealer, on both synthetic scenarios and on realistic epidemic scenarios calibrated on the COVID-19 spread in Italy. Our results prove both the effectiveness of the quantum-based approach (which is typically comparable, or even superior to other methods) and its computational efficiency, concerning which it outperforms other classical methods.

The preliminary results presented in this work open several promising directions for future research. First, our method was developed under the assumption of a fixed two-step optimization time window. Future work could explore the ability of quantum solvers to handle longer time horizons. In particular, a comparison could be conducted between results obtained using QAOA --- which can directly solve the PUBO problem --- and those from quantum annealing (QA), which requires a polynomial reduction to QUBO that may introduce auxiliary variables. To enable the analysis of larger networks, QUBO preprocessing techniques can be explored~\cite{orlandi2024qoolchain}. Additionally, quantum-inspired solvers~\cite{volpe2023integration, volpe2024improving} could be investigated as a mature and scalable alternative to purely quantum approaches.
 Furthermore, while the present study focuses on control actions based on restricting mobility between nodes, other types of interventions --- both pharmaceutical and non-pharmaceutical --- can be considered in epidemic response strategies. Extending our approach to accommodate and compare different intervention types, in terms of both healthcare impact and social and economic cost, represents a key avenue for further investigation. Finally, quantum computing has demonstrated effectiveness in solving optimization problems within network-based epidemic models, suggesting that similar approaches could be applied to other domains of complex network control, such as influence maximization in social networks or formation control in multi-agent systems.

\section*{acknowledgements}
This research was supported by the Department of Electronics and Telecommunications, Politecnico di Torino, under the project \textit{ALLOY: Quantum Control of Complex Network Systems}. The authors thank CINECA for the collaboration and access to dwave machines through the ISCRA program.
\bibliographystyle{ieeetr}
\bibliography{sn-bibliography}% common bib file

\end{document}

%% file: img/network.tex
\begin{tikzpicture}
\definecolor{mygreen}{RGB}{72,160,66}
\definecolor{myyellow}{RGB}{190,170,0}
\definecolor{myviolet}{RGB}{195,8,255}
\definecolor{mylightblue}{RGB}{173, 216, 250}

\foreach \x/\y/\n in {.1/0/1,2.1/-.25/2,1.5/1.5/3,3.2/.8/4,-1.5/0.5/5,-3.8/.75/6,-2.7/2/7,-.2/2.2/8}
{\node[draw=mylightblue!60,  inner sep=0pt, circle, minimum size=18pt,fill=mylightblue!60] (\n) at  (\x,\y) {\small{\textcolor{black}{$\n$}}};}

\tikzset{mystyle/.style={-,gray,thick,font=\small}} 
\tikzset{every node/.style={fill=white}}

\foreach  \i/\j/\w in {6/7/0.3,5/1/0.2,8/5/0.4,8/3/0.3,7/8/0.1,7/5/0.2,3/1/0.1,3/4/0.4,4/2/0.1,1/2/0.15,6/5/0.05}
{\path (\i) edge[mystyle] node {$\w$} (\j) ;}

\end{tikzpicture}

%% file: img/sis.tex
\definecolor{mygreen}{RGB}{72,160,66}
\definecolor{myyellow}{RGB}{190,170,0}
\definecolor{myviolet}{RGB}{195,8,255}
\definecolor{mylightblue}{RGB}{173, 216, 250}
\definecolor{myred}{RGB}{200, 0, 0}
\begin{tikzpicture}
\node[draw=mylightblue, fill=mylightblue!10,circle, very thick,minimum size=.6cm] (S) at (0,0) {$S$};
\node[draw=myred, fill=myred!10,circle, very thick,minimum size=.6cm]  (I) at (2.2,0) {$I$};
\path [->,>=latex,ultra thick]  (S) edge [bend left=30]  node [above] {{$\lambda$}} (I);
\path [->,>=latex,ultra thick]  (I) edge[bend left=30]   node [below] {{$\mu$}} (S);
\node[draw=none] (S) at (0,-1) {};
\end{tikzpicture}

%% file: img/sir.tex
\definecolor{mygreen}{RGB}{72,160,66}
\definecolor{myyellow}{RGB}{190,170,0}
\definecolor{myviolet}{RGB}{195,8,255}
\definecolor{mylightblue}{RGB}{173, 216, 250}
\definecolor{myred}{RGB}{200, 0, 0}
\begin{tikzpicture}
\node[draw=mylightblue, fill=mylightblue!10,circle, very thick,minimum size=.6cm] (S) at (0,0) {$S$};
\node[draw=myred, fill=myred!10,circle, very thick,minimum size=.6cm]  (I) at (2,0) {$I$};
\node[draw=myviolet, fill=myviolet!10,circle, very thick,minimum size=.6cm]  (R) at (4,0) {$R$};
\path [->,>=latex,ultra thick]  (S) edge  node [above] {{$\lambda$}} (I);
\path [->,>=latex,ultra thick]  (I) edge   node [above] {{$\mu$}} (R);
\node[draw=none] (S) at (0,-1) {};
\end{tikzpicture}